\documentclass{aa}  

\usepackage{graphicx}
\usepackage{epstopdf} 
\usepackage{txfonts}
\usepackage{amsmath}
\usepackage{natbib}
\usepackage{color}

\bibpunct{(}{)}{;}{a}{}{,} 

\begin{document}

   \title{The origin of quasi-periodicities during circular ribbon flares}


   \author{L. K. Kashapova
          \inst{1}
          \and
               E. G. Kupriyanova
          \inst{2}
          \and
               Z. Xu
          \inst{3}
          \and
              H. A. S. Reid
          \inst{4}
          \and
              D. Y. Kolotkov
          \inst{5, 1}
          }

   \institute{Institute of Solar-Terrestrial Physics SB RAS, Lermontova st. 126A, 664033 Irkutsk, Russia\\
              \email{lkk@iszf.irk.ru}
         \and
                 Central (Pulkovo) Astronomical Observatory of the RAS, 196140 Saint Petersburg, Russia\\
              \email{elenku@bk.ru}
         \and
                 Yunnan Observatories, Chinese Academy of Sciences,  Kunming, Yunnan, China \\
              \email{xuzhi@ynao.ac.cn}
         \and
                 School of Physics and Astronomy, University of Glasgow, Glasgow G12 8QQ, UK\\
                 Department of Space and Climate Physics, University College, London, RH5 6NT, UK \\
             \email{hamish.reid@ucl.ac.uk}
         \and
                 Centre for Fusion, Space and Astrophysics, Department of Physics, University of Warwick, CV4 7AL, UK\\
              \email{d.kolotkov.1@warwick.ac.uk}
             }

   \date{Received ; accepted }

  \abstract
{Solar flares with a fan-spine magnetic topology can form circular ribbons. The previous study based on H$\alpha$ line observations of the solar flares during March 05, 2014 by \cite{Xu2017ApJ} revealed uniform and continuous rotation of the magnetic fan-spine. Preliminary analysis of the flare time profiles revealed quasi-periodic pulsations (QPPs) with similar properties in hard X-rays, H$\alpha$, and microwaves.}
{In this work, we address which process the observed periodicities are related to: periodic acceleration of electrons or plasma heating?}
{QPPs are analysed in the H$\alpha$ emission from the centre of the fan (inner ribbon R1), a circular ribbon (R2), a remote source (R3), and an elongated ribbon (R4) located between R2 and R3. The methods of correlation, Fourier, wavelet, and empirical mode decomposition are used. QPPs in H$\alpha$ emission are compared with those in microwave and X-ray emission.}
{We found multi-wavelength QPPs with periods around 150~s, 125~s, and 190~s. The 150-s period is seen to co-exist in H$\alpha$, hard X-rays, and microwave emissions, that allowed us to connect it with flare kernels R1 and R2. These kernels spatially coincide with the site of the primary flare energy release. The 125-s period is found in the H$\alpha$ emission of the elongated ribbon R4 and the microwave emission at 5.7~GHz during the decay phase. The  190-s period is present in the emission during all flare phases in the H$\alpha$ emission of both the remote source R3 and the elongated ribbon R4, in soft X-rays, and microwaves at 4--8~GHz.}
{We connected the dominant 150-s QPPs with the slipping reconnection mechanism occurring in the fan. We suggested that the period of 125~s in the elongated ribbon can be caused by a kink oscillation of the outer spine connecting the primary reconnection site with the remote footpoint. The period of 190~s is associated with the 3-min sunspot oscillations.}

   \keywords{Sun: flares; Acceleration of particles; Magnetic reconnection; Sun: radio radiation; Sun: X-rays, gamma rays; Sun: chromosphere; Sun: corona}

   \maketitle

\section{Introduction}\label{s:Introduction}
In contrast to standard two-ribbon solar flares \citep{1964NASSP..50..451C, 1966Natur.211..695S, 1974SoPh...34..323H, 1976SoPh...50...85K}, circular-ribbon flares are accompanied by brightenings of chromospheric emission in a quasi-circular manner. The simplest model of the magnetic field for these flares (so-called fan-spine configuration) was proposed by \citet{2009ApJ...691...61P}. The structure consists of a dipolar magnetic field, embedded into a dome-structured magnetic field. The principal structural elements are fan lines with an inner and outer spine \citep[see, for example, Figure~1 in][]{2009ApJ...691...61P}.
The cross-section of the fan lines and the chromosphere plots a quasi-circle. The inner spine draws the inner ribbon (inside the circular ribbon). The outer spine, originating from the dome-like structure, can form open magnetic field lines giving the possibility that energetic particles travel outwards from the corona to interplanetary space.  Alternately, the outer spine can be anchored at a remote site in the chromosphere \citep{2000ApJ...540.1126A, 2009PASJ...61..791M, 2015ApJ...804....4K}. In  this case, the outer spine is associated with a loop visible in soft X-rays \citep{2001ApJ...554..451F} or in chromospheric line emission \citep{2015ApJ...804....4K}. The brightening  of the circular ribbon is accompanied with the brightening of its remote footpoint \citep{2009PASJ...61..791M, 2015ApJ...804....4K, Xu2017ApJ}. 

The various explanations of the mechanisms of energy release in this type of flare are based on the presence of a quasi-separatrix layer with the magnetic null-point between the fan and dome. 
In this model of magnetic field configuration, the sheared magnetic field of the sunspots
prompts reconnection at the magnetic null-point \citep{1990ApJ...350..672L, 2000ApJ...540.1126A, 2009ApJ...704..341S}. 
Observational confirmation of this hypothesis was revealed  by \citet{2009ApJ...700..559M}. Based on EUV observations, they found the anticlockwise running brightening of the fan magnetic lines.  Such brightenings can be explained by slipping or slip-running reconnection.  There is evidence by several authors \citep[see, for example,][]{2009PASJ...61..791M,Ried2012A&A,2015ApJ...805....4Z}  for detecting  the hard X-ray emission with energy above 25 keV in circular-ribbon flares that was interpreted   as bremsstrahlung emission  by  non-thermal electrons. 

The processes of particle acceleration and heated plasma motion are usually going on simultaneously in a solar flare. The structure of the circular ribbon flare defined by the magnetic fan and null point indicates the most probable place of acceleration and primary energy release. 
Moreover, in contrast to two-ribbon flare, the location of this place within the circular-ribbon structure does not move during the flare \citep{2009PASJ...61..791M,Xu2017ApJ}. 
From this point of view, the circular ribbon flare is a more convenient object for the study of acceleration and energy propagation processes then the classical two-ribbon flare. Analysis of quasi-periodic pulsations (QPPs) in circular ribbon flares can be helpful for diagnostics of wave processes or particle acceleration as pulsation generators. Despite a number of studies of circular-ribbon flares, the manifestations of QPPs in this type of flare has rarely been revealed and analysed.
In spite of the similar magnetic configurations of circular-ribbon flares where QPPs were observed, the results of these studies indicate the various reasons for their onset. \citet[][]{2013SoPh..283..473M} proposed that  QPP  with  periods of 10--83~s observed in radio wavelengths are connected with fast magnetoacoustic waves which propagate along the fan outwards from the magnetic null point.
\citet[][]{2015ApJ...804....4K} found a global standing slow magnetoacoustic wave with a period of 409~s in the hot channels (94 and 131~{\AA}) registered with \textit{the Atmospheric Imaging Assembly} onboard \textit{the Solar Dynamics Observatory} (SDO/AIA). The wave was reflected several times between the footpoints of the outer spine. Simultaneously, a period of 202~s was detected in the 6--12~keV X-ray emission  in the remote footpoint of outer spine. This periodicity was associated by the authors with possible repetitive reconnection in the fan that caused an acceleration of electrons. 
The results were obtained in different spectral ranges that correspond to different layers of the solar atmosphere and different mechanisms of emission excitation. We guess that one of the reasons for such heterogeneous results could be that the viewing angle allowed for detecting emission from only one of the structures composing the fan-spine structure of the circular ribbon flare.

We need to find an event such that the angle of view allows us to study each part of the fan-spine structure independently. Moreover an event should be observed simultaneously at different ranges of the electromagnetic spectrum covering solar atmosphere heights from the chromosphere to the upper layers of the corona. 
The fan-spine structure located near to the solar disk centre produced a series of consecutive flares suitable for all these criteria during  March 5,  2014.  We chose for analysis the event presented by~\cite{Xu2017ApJ}. The event consisted of two flares (C2.8 and M1.0 by GOES (\textit{Geostationary Operational Environmental Satellite}) classification) with well-pronounced QPPs seen not only in H$\alpha$ emission but at the other wavelengths.
The purpose of our study is to check if the QPPs observed in the chromospheric H$\alpha$ emission have the same nature as the QPP in hard X-ray and microwave emissions. We also analyse a connection between the QPP at the fan structure with the QPP at the remote source in order to understand if they have the common driver of pulsations or not.

The paper is organised as follows. The observations are described in Section~\ref{s:Observations}. The spatial-temporal features of the flare seen in the multi-wavelength observations are summarised in Section~\ref{s:Spatial}. The analysis of flare time profiles is carried out in Section~\ref{s:Timeprofs}. The detected periodicities are analysed in Section~\ref{s:Periods}. The mechanisms are discussed and results are summarised in Section~\ref{s:Discussion}. Details of the data processing technique are described in Appendix~\ref{s:Method}.

\section{Observations}\label{s:Observations}
Our analysis is based on observations of  the chromospheric emission in H$\alpha$ line obtained by \textit{the New Vacuum Solar Telescope} (NVST) \citep{2014RAA_NVST}. The telescope is equipped with the Lyot filter system with the spectral band-pass in 0.25~{\AA}. The images are obtained  with 0.5 arcsec spatial resolution and a 12 second  time resolution.  We used the data for analysis of the chromospheric response to precipitation of energetic particles and gas-dynamic processes like heating.  We take information about the solar flaring plasma behaviour from microwave and X-ray observations.

Microwave flare emission was obtained using data from \textit{the Siberian Solar Radio Spectropolarimeter} (SSRS), the former \emph{Badary Broadband Microwave Spectropolarimeter}  \citep{SSRS2011CEAB,SSRS2015SoPh}. SSRS observes  the solar microwave fluxes  at 16 frequencies within the 3.8--8.2 GHz range. The observations are carried out with a cadence of 0.6~s.  
We found a very faint response in the correlation plots at 17~GHz obtained by \textit{the Nobeyama Radioheliograph} \citep{Nakajima_etal_1994} but  the registered flux is too low to get any valid information about the spatial structure of the microwave source.

We use the data from three different instruments for the analysis of the X-ray emission. \textit{The Reuven Ramaty High Energy Solar Spectroscopic Imager} (RHESSI) \citep{Lin:2002aa} observed both flares but we cannot use its time profiles for the analysis of  QPPs because of  several switches between attenuators during the observations. However, the RHESSI data are appropriate to reconstruct the images within the time intervals between the switches.  We reconstructed the images at the energies 6--12~keV, 12--25~keV and 25--50~keV. 
In some time intervals,
the numbers of counts were not enough to do correct imaging for all set of energies. We also use the X-ray spectra by RHESSI for estimation of the flare plasma parameters. 
\textit{The Gamma-Ray Burst Monitor}  (hereinafter, Fermi/GBM) of \textit{the Fermi Gamma-ray Space Telescope} \citep{Meegan2009ApJ} did not observe the full time interval of interest. The advantage of this instrument is the absence of instrumental artefacts affecting periodicity detection. Thus we used Fermi/GBM data for proofing solar origin of the signal.
The observations both by RHESSI and Fermi/GBM were processed by the standard software \citep{2002SoPh..210..165S}. The X-ray time profiles suitable for the periodicity analysis are obtained by \textit{Konus, a Gamma-Ray Burst Experiment} \citep{Aptekar1995,Palshin_etal_2014} on-board \textit{the Wind} spacecraft (hereinafter, KW)  in  the energy band 20--80~keV. The temporal resolution of  these data is 2.944~s.
Soft X-ray emission was analysed using GOES within two bands: 1--8~{\AA} and 0.5--4~{\AA}. The time resolution of this data is  2.05~s.

We use the data of SDO to analyse both the line-of-sight magnetograms obtained with \textit{the Helioseismic and Magnetic Imager} (SDO/HMI)  \citep{HMI2012SoPh} and the images obtained by SDO/AIA \citep{Lemen:2012aa}.

\section{Spatial structure and temporal evolution of the flare sources}\label{s:Spatial}
\begin{figure*}
	\centerline{
		\includegraphics[width=0.75\textwidth,clip=]{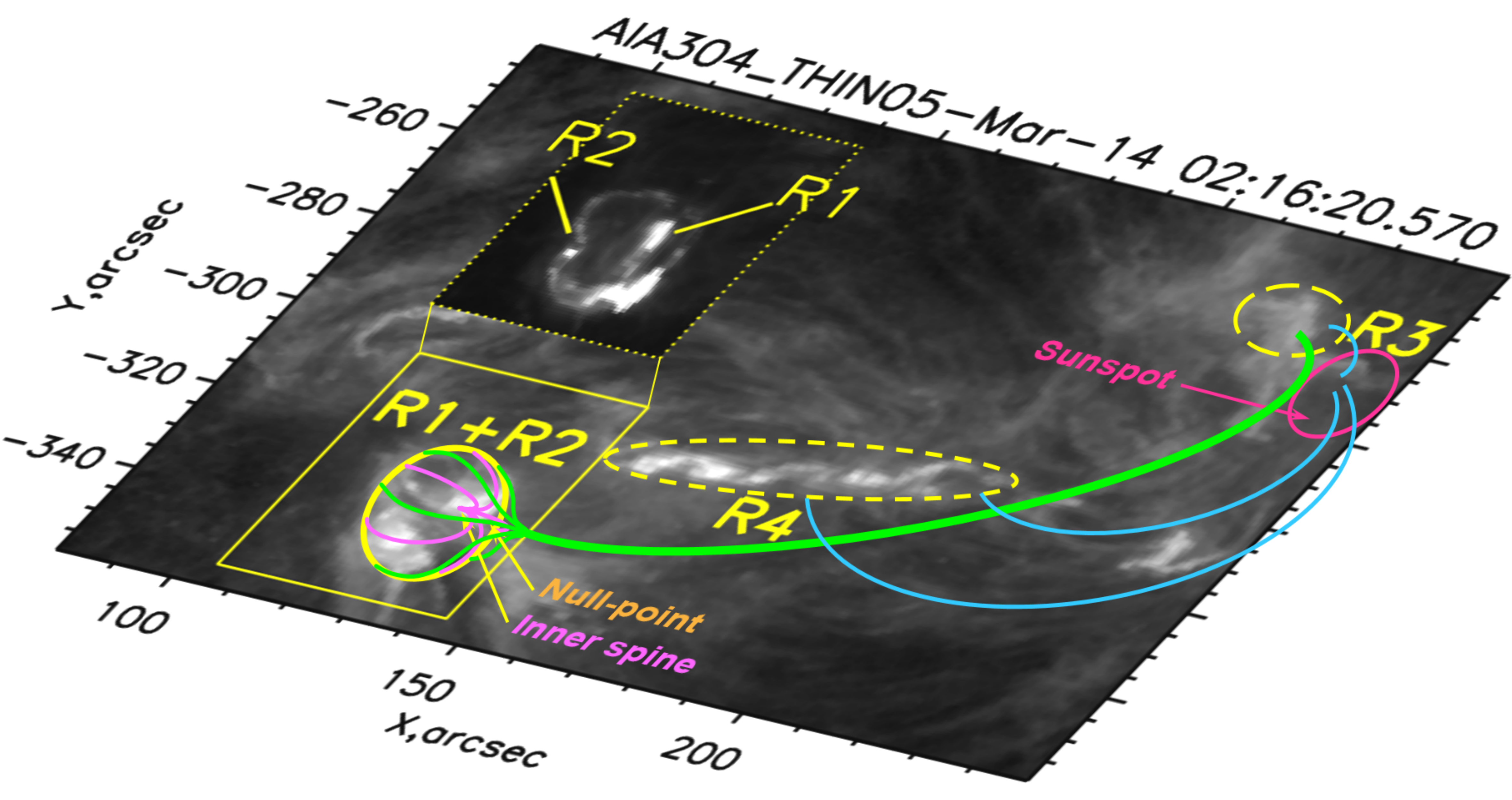}}
	\caption{The principal elements of the fan-spine structure involved in the flare SOL2005-05-06 schematically sketched over the SDO/AIA image in the 304~{\AA} band at 02:16:20~UT. The chromospheric kernels are outlined by the yellow ovals: R1+R2 is the circular ribbon structure, where the inner ribbon R1 and the circular ribbon R2 are shown in detail separately at 01:55:49~UT within the dotted rectangle; R3 is the remote ribbon; R4 is the elongated ribbon seen at the decay phase. The fan-spine magnetic  structure is drawn in green and magenta: the fan lines (magenta) form the inner spine; the dome lines (green) continue into the outer spine (a long loop anchored at R3 kernel); a quasi-separatrix layer is the surface between magenta and green lines (is not shown here) with the magnetic null-point located between the inner and outer spines \citep{2009ApJ...691...61P}. Loops in blue connect the sunspot (dark pink) with R3 and R4 kernels. }
	\label{f:sketch}
\end{figure*} 
\begin{figure*}
   \centerline{
\includegraphics[width=0.7\textwidth,clip=]{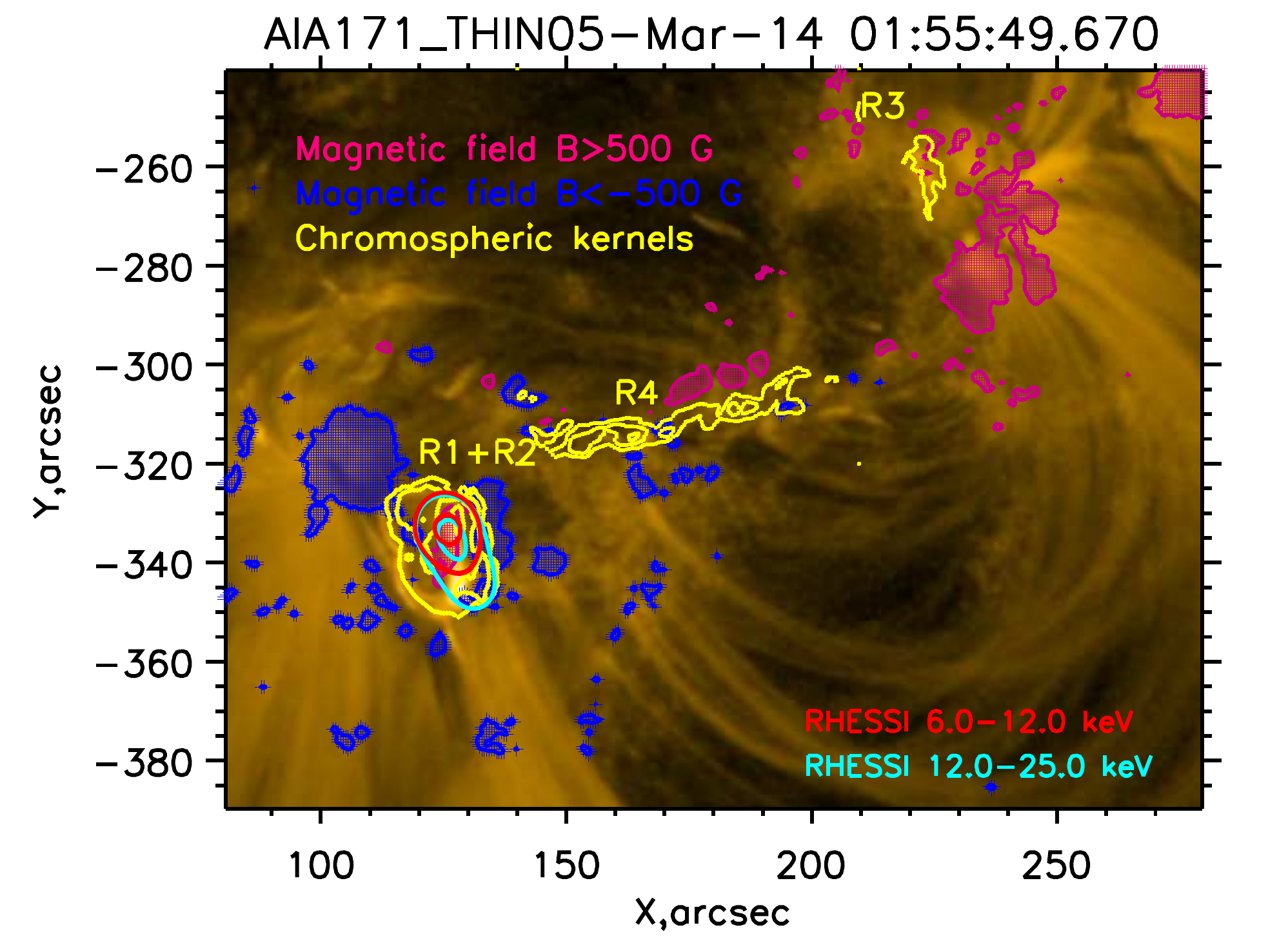}
}
   \centerline{
\includegraphics[width=0.7\textwidth,clip=]{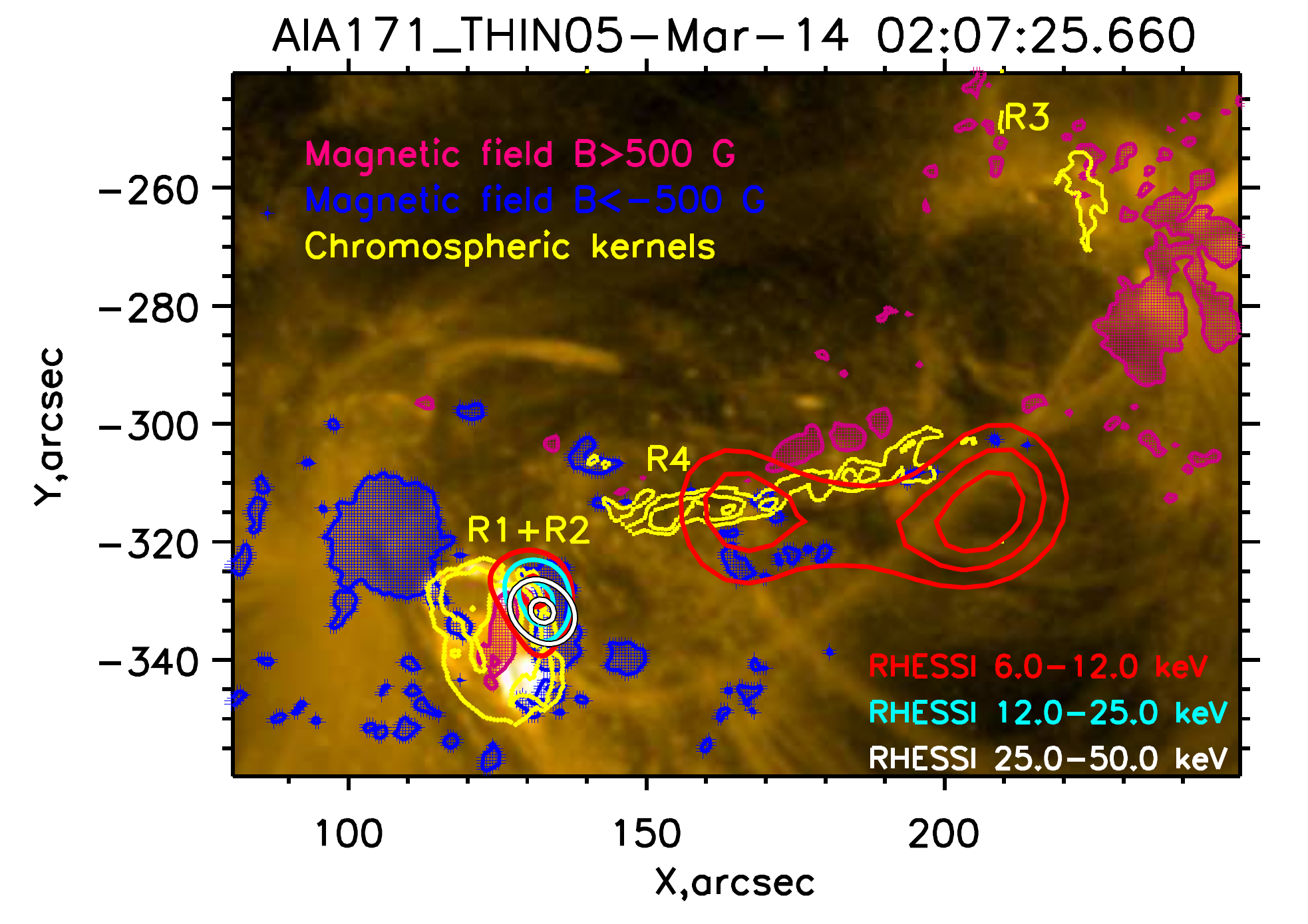}
  }
\caption{SDO/AIA 171~{\AA} maps during the first flare at 01:57:12~UT (top) and during the second flare at 02:07:25~UT (bottom) with the over-plotted contours of hard X-ray emission at 6--12~keV (red), 12--25~keV (light green) and 25--50~keV (white). In the top panel, the X-ray images were reconstructed within 01:55:15--01:55:35 UT (the first peak) by the CLEAN algorithm. The X-ray sources shown by the solid line in the bottom panel were reconstructed within 02:07:00--02:08:40 UT by PIXON algorithm. The contour levels are 50\% and 90\% of maximum. The X-ray source associated with R4 was obtained by CLEAN algorithm within 02:18:30--02:20:30~UT. The levels are 40\%, 60\% and 80\% of maximum.
The SDO/AIA 304~{\AA} chromosperic kernels are marked by yellow. The magnetic field features of positive and negative polarity are marked by blue and magenta, correspondingly. The absolute value of magnetic feature strength is above 500~G.}
  \label{f:SDO}
\end{figure*} 
The circular-ribbon flares  occurred in the active region NOAA~11991 close to the solar disk centre ($X = 120$, $Y = -340$ arc seconds) on March 5, 2014 (Figure~\ref{f:sketch}).  
\citet{Xu2017ApJ} carried out analysis of the flare topology based on  $H\alpha$  observations and  SDO/HMI magnetorgams. They revealed that the principal elements of the fan-spine structure are well mapped in $H\alpha$ emission \citep[see Figure 1 in ][]{Xu2017ApJ}. 
In Figure~\ref{f:sketch}, we draw the fan-spine structure over the chromospheric image in the 304~{\AA} SDO/AIA channel which also demonstrates the same form of the flare ribbons as $H\alpha$ observations. Hereafter, we used the same labels for the flare ribbons/kernels as \citet{Xu2017ApJ}. Particularly,
the inner ribbon/kernel  (R1) corresponds to the inner spine. The footpoints of the fan
and dome
draw the circular ribbon (R2). 
Note that with time these kernels get strongly entangled so that it becomes impossible to separate their fluxes correctly every time \citep{Xu2017ApJ}. Therefore, in this paper we analyse the whole system R1+R1. 
The remote footpoint of the outer spine is marked by R3. 
The authors also detected the ribbon, marked by R4  which appeared later than the other ribbons and after the main phase of the event.  Hereafter, we used the same labels for the flare ribbons/kernels as \citet{Xu2017ApJ}.
The circular ribbon appeared on March 4, 2014 at 08:00~UT and existed till approximately  05:00~UT on March 5, 2014.  
The comparison of the magnetogram  and the location of the chromospheric flare kernels (ribbons R1$+$R2) shows that the circular ribbon coincides with the magnetic feature of positive polarity which is surrounded by the features of the negative magnetic polarity (Figure~\ref{f:SDO}).  The remote kernel R3 is connected by the loop to the fan structure R1 and R2.
This loop corresponded to the outer spine and is seen in the 171~{\AA} SDO/AIA channel image after the maximum of the second flare  (bottom panel of Figure~\ref{f:SDO}). 

The time profile of the $H\alpha$ emission during interval 01:50--02:40~UT  \citep{Xu2017ApJ} could be identified as a set of quasi-periodic pulses. GOES classified them as two events. The first event was a C2.8 GOES class flare  with the onset  at 01:52~UT and the maximum at 01:58~UT. The second event was an M1.0 GOES class flare with the onset at 02:06~UT and the maximum at 02:10~UT. However, the time profiles in other wavelengths demonstrate a series of peaks between and outside the GOES flares (see Figure~\ref{f:timeprofs}) indicating that they are parts of a common process. Therefore, despite the GOES classification, we study these peaks as a single continuous event.

We reconstructed the X-rays images  with the CLEAN and PIXON algorithms using the package from Solar Software \citep{2002SoPh..210..165S} based on data of the 4--8 front detectors.
The centre of the X-ray sources at  6--12~keV and at 12--25~keV bands coincide with the inner ribbon R1 during 
the first flare peak
(Figure~\ref{f:SDO}, the top panel).  However the 12--25~keV source also covers part of the ribbon R2.  In spite of some flux being detected at energies above 25~keV, we could not reconstruct source images for these energies.
Later, 
there are X-ray sources up to 50~keV (Figure~\ref{f:SDO}, the bottom panel). 
Compared with its earlier location,
the sources moved towards the circular ribbon R2 and localise at one specific part.
In spite of the outer-spine loop which connects the fan region (R1$+$R2) with the remote source (R3), we did not reveal any X-ray source coinciding with the remote source R3. However, it could be too faint to be distinguished near to the bright source (R1$+$R2).

As mentioned above, the $H\alpha$ emission from the inner ribbon R1 dominates during the impulsive phases 
of the analysed event
(Figure~\ref{f:timeprofs}).  The exception is the interval between 02:02~UT and 02:07~UT when the maximal emission is produced by the southern chord of circular ribbon R2.  
The peaks in the time profile of the remote source R3 (red curve) show a delay relative to peaks of the source R1 (blue curve).

The brightening of the elongated ribbon R4 (red curve) occurs at the final phase of the flare activity. However, the variations of the R4 time profile showed a sharp rise of the flux and a complicated structure. One can see in Figure~\ref{f:SDO} (bottom panel) that kernel R4 is associated with 6--12~keV X-ray sources. That could be a result of a small flare loop arcade appearance.
\section{Analysis of multi-wavelengths spectra and time profiles }\label{s:Timeprofs}
\begin{figure*}
	\centerline{
		\includegraphics[width=0.7\textwidth,clip=]{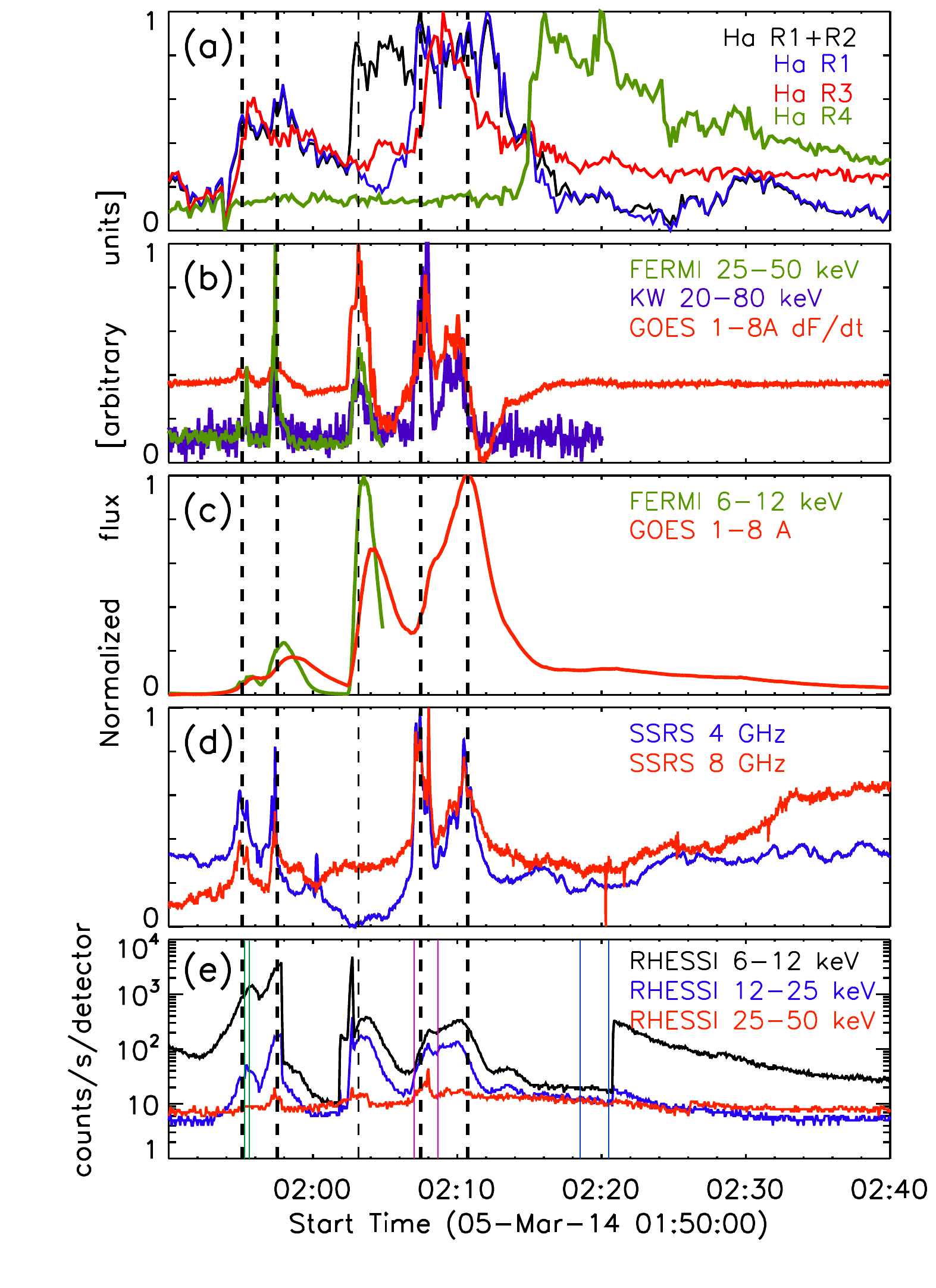}
	}
	\caption{
		Normalised time profiles of the flare emission in different spectral ranges. All the time profiles are normalised by their maxima. A colour of each curve within a panel corresponds to the colour of the legend in the panel. Panel (a): H$\alpha$ emission from the sources R1+R2, R1, R3, and R4. Panel (b): X-ray flux observed by Fermi/GBM at 25--50~keV and KW at 20--80~keV, and derivative of the GOES flux, at 1--8~{\AA}. Panel (c): X-ray fluxes registered by Fermi/GBM 6--12 keV, by GOES at 1--8~{\AA}, and by GOES at 0.5--4~{\AA}. Panel (d): microwave fluxes at 8~GHz, 5.7~GHz, and 4~GHz obtained with SSRS. Panel (e): X-ray uncorrected fluxes registered by RHESSI at 6--12~keV, 12--25~keV, and 25--50~keV. The RHESSI sources shown in Figure~\ref{f:SDO} were reconstructed within time intervals enclosed into pairs of the same-coloured vertical solid lines.
	}
	\label{f:timeprofs}%
\end{figure*} 
\begin{figure*}
	\centerline{
		\includegraphics[width=0.5\textwidth,clip=]{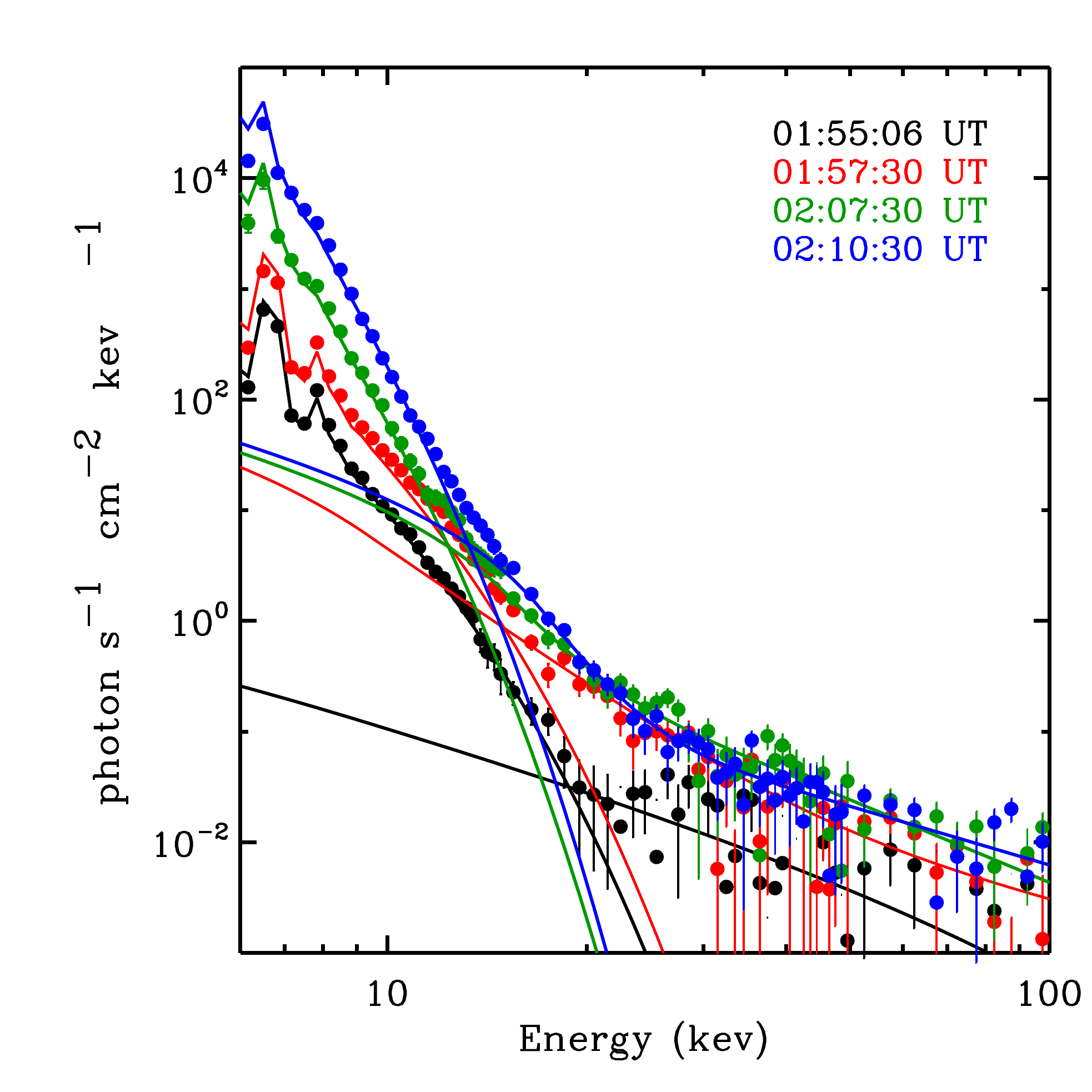}
		\includegraphics[width=0.485\textwidth,clip=]{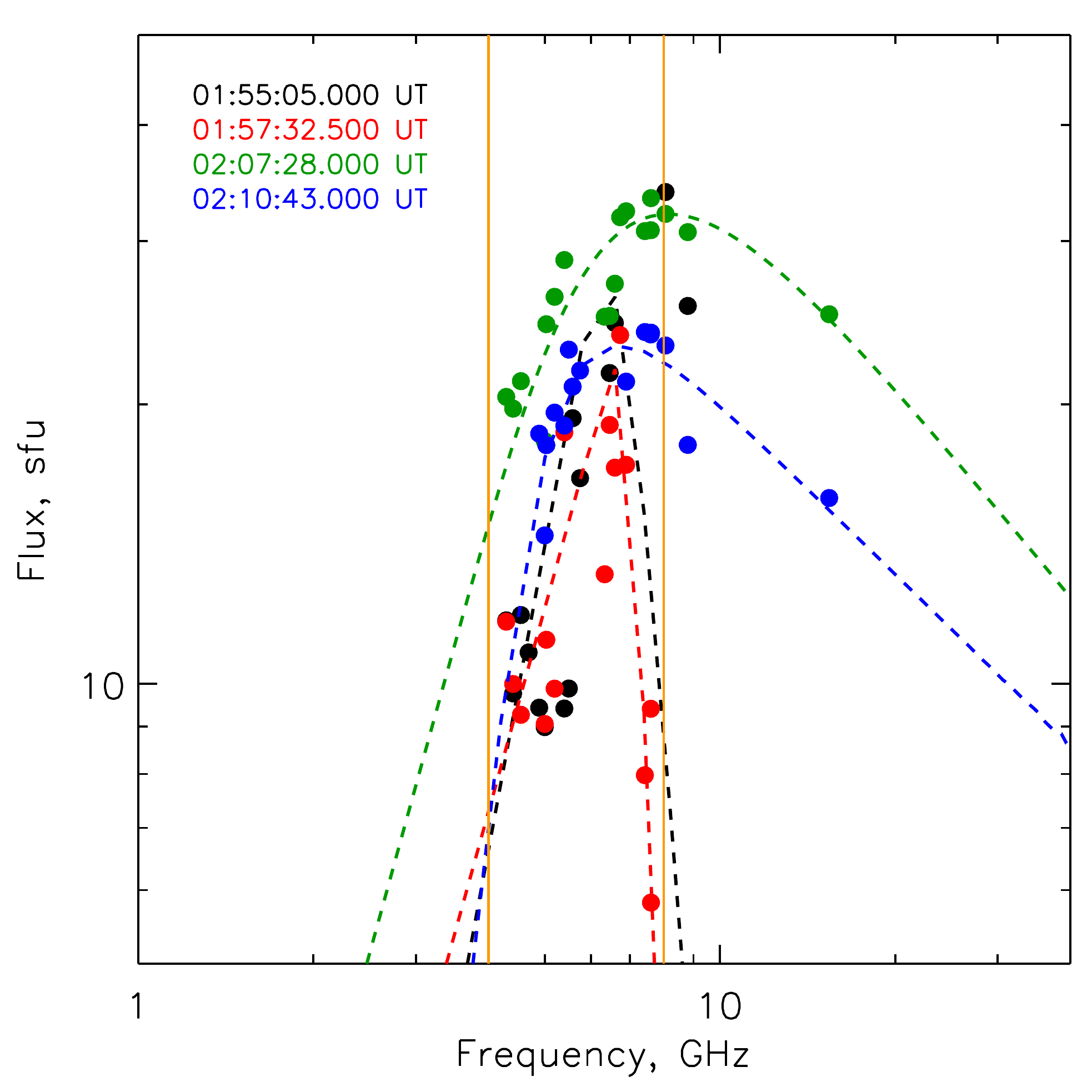}
	}
	\caption{
		Left panel: evolution of the X-ray spectra.  The observed X-ray spectra (marked by dots) are over plotted by the results of the fitting by a model consisting of an optically thin thermal bremsstrahlung
		function  and power-law function. The average value of the time interval, the observational data and the results of the fitting are marked by the same colour for each moment in time.
		Right panel: evolution of the microwave spectra.
		The observed data (marked by the dots) are over plotted by the result of fitting by the function corresponding to the form of the gyrosynchrotron spectrum.  The average value of the time interval, the observational data and the results of the fitting are marked by the same colour for each moment in time.  The vertical orange lines mark the 4 GHz and 8 GHz frequencies.}
	\label{f:Spectra}%
\end{figure*} 

The considered events demonstrated a wide range of electromagnetic emission~--- from optical chromospheric emission to soft and hard X-rays  and microwaves. 
H$\alpha$ emission allows us to study relationships between QPPs present  in the spatially resolved emission from different flare kernels. However, it is not possible to identify the mechanism responsible for the QPPs using only H$\alpha$ data. This emission could be related to both accelerated electrons (non-thermal mechanisms) and heating of the thermal plasma (thermal mechanisms). We need to distinguish the emission mechanisms first using spectral analysis of  X-rays and microwave (MW) data. We can use the similarity of the QPPs observed in X-ray and microwave emissions and those in the H$\alpha$ emission as a marker of the specific mechanism in the kernel. This section is devoted to analysis of spectra of the X-ray and microwave emissions and to analysis of the relationships between time profiles of different emissions.
 
Figure~\ref{f:Spectra} (left panel) shows the spectra obtained from RHESSI data for the four main peaks marked with dashed vertical lines in Figure~\ref{f:timeprofs}. These peaks are clearly seen in X-rays, microwaves and H$\alpha$. 
We fitted the spectra by two models with albedo correction because of the event location near to the solar disk centre. The first model was simpler and consisted of a thermal bremsstrahlung function and a power-law function (non-thermal emission). Using this model, we got the electron temperature varying from 17~MK to 14~MK during the flares. 
The second model consisted of multi-thermal bremsstrahlung function where the differential emission measure has a power-law dependence on temperature and thick target bremsstrahlung function. The maximal plasma temperature for this model varied from 23~MK to 14~MK. The estimations of the plasma temperature are similar for both models.
The high energy cut-off of the thermal component was not higher than 20~keV during the flares for both models. Thus, we can assume that all X-ray emission above 20~keV is non-thermal, and we can use it as an indicator of electron acceleration processes.

The main peaks in both the microwave and KW hard X-ray emission
correlate with the peaks in the time derivatives of the GOES flux manifesting the Neupert effect \citep{1968ApJ59N} applicable to the case.
The exception was the third X-ray peak at 02:03~UT, approximately, which does not have a strong response in the microwaves. The electron flux is low but still high enough for the generation of gyrosynchrotron emission. We infer that a lack of significant MW emission results from the combination of low magnetic field values and the soft, small flux of accelerated electrons.  The final answer could be obtained through modelling of the MW emission, taking into account the peculiarity of the magnetic field topology.  However, this is beyond the scope of our current study.  All other peaks
show peak-to-peak agreement with those peaks seen in the H$\alpha$ time profiles  of the R1+R2 source and R1 source. H$\alpha$ line centre emission could be both a result of direct ionization of chromospheric plasma by beams of accelerated electrons, and/or plasma heating.  The heating could be the result of gas-dynamic expansion following excitation by non-thermal particles or the result of other thermal energy release processes. Thus, the peaks seen in the H$\alpha$ line profiles could have different origins. In this case, there would be two or more dominating periods in the time profiles of chromosperic emission. Even if we would find a single period  accompanied by high noise background, then we will have several independent models supporting both the thermal and non-thermal nature of the QPPs. That is why we have to use the periodicity obtained from the time profiles of non-thermal X-rays as a marker of the acceleration processes.

Figure~\ref{f:timeprofs} (panel (c)) shows a correlation between the H$\alpha$ line profiles of the R1+R2 and R1 sources and thermal X-ray emission. According to the spectral analysis (Figure~\ref{f:Spectra}), the X-ray emission is thermal for energies below about 20~keV. However observations of Fermi/GBM stopped about 02:05 UT and we have to use X-ray data by GOES for testing thermal emission of the flares. The GOES band 1--8~{\AA}  corresponds to about 1.5--12~keV energy band  while emission of the GOES band 0.5--4~{\AA} \ is equal to  3--25~keV energy band. The 1--8~{\AA} GOES time profile shows a good correlation with the 6--12~keV time profile obtained by Fermi/GBM. All peaks of GOES time profile correlate with those peak corresponding to H$\alpha$ kernels. The peaks of the 1--8~{\AA} GOES  time profile derivative also demonstrate the good correlation with non-thermal X-ray emission obtained by Fermi/GBM and KW. Thus we use both 1--8~{\AA} band time profile and its derivative as indicators of thermal and non-thermal processes in the discussed event.

Microwave spectra for the same time periods as the X-ray emission are shown in right panel of Figure~\ref{f:Spectra}. We see that the peak frequency is about 6 GHz and does not significantly shift towards higher frequencies during the evolution of the 
analysed event.
Thus, the flux at 8 GHz locates optically thin microwave emission and is a direct marker of the emission generated by accelerated electrons. The flux at 4 GHz will be an indicator of the processes generating emission in the optically thick part of the spectrum.  Both frequencies are marked by vertical orange lines in the Figure~\ref{f:Spectra} (right panel).

The peaks seen in the H$\alpha$ time profiles of the R1+R2 sources together and the R1 source alone (Figure~\ref{f:timeprofs}, panel (a)) 
appeared at the same time as the main peaks 
in the hard X-ray  (Figure~\ref{f:timeprofs}, panel (b)) and microwave emissions at the selected frequencies (Figure~\ref{f:timeprofs}, panel (d)). 
The time profiles of microwave fluxes at frequencies above and below the peak frequency show good agreement with each other. We note that time profiles of the microwave fluxes at frequencies above and below the peak frequency show in-phase variations. This fact means that both the optically thin part and optically thick part of the microwave spectrum indicate the same process of electron acceleration without any significant impact of self-absorption by plasma in the flare volume. In other words, both the optically thin and optically thick microwave emission is caused mainly by gyrosynchrotron emission from the same population of accelerated electrons. 

Summarising the results presented in the current section, the spectral analysis allowed us to separate emission from thermal and non-thermal generation mechanisms. The similar behaviour of the microwave, non-thermal X-ray, and chromospheric emission gives us evidence that all  flux variations have a solar origin and are not artefacts. Thus, these time profiles could be used for periodicity testing and obtaining periodograms for the different H$\alpha$ sources and different spectral bands. Comparative analysis of periodograms obtained for these emissions and  H$\alpha$ emission will help us to reveal which processes dominate in emission generation of the different flare sources. Moreover, information about periodicity of the different H$\alpha$ sources will address the question about the existence of a general process controlling the emission of all flare sources during both events. 

\section{Periodic properties of the flares}\label{s:Periods}
\begin{figure*}
	\centerline{
		\includegraphics[width=0.45\textwidth,clip=]{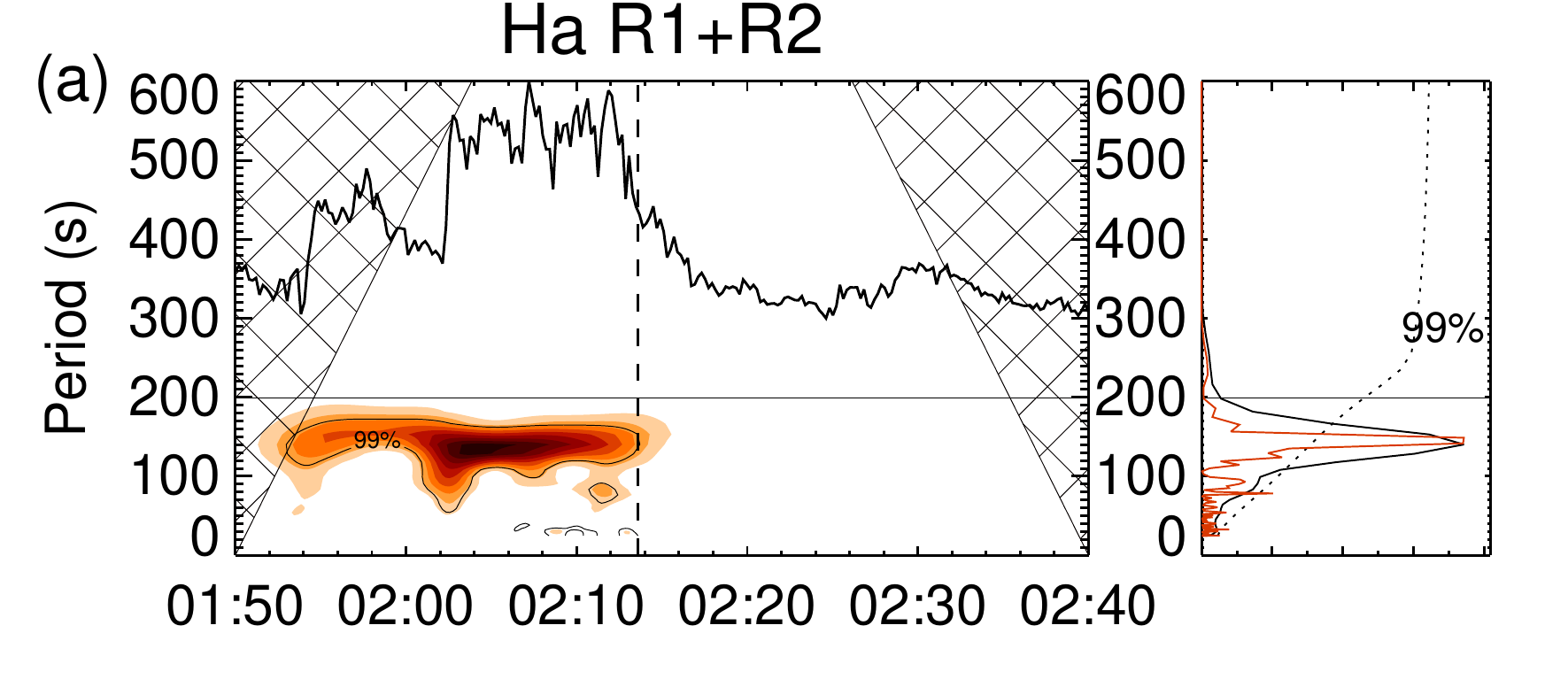}
		\includegraphics[width=0.45\textwidth,clip=]{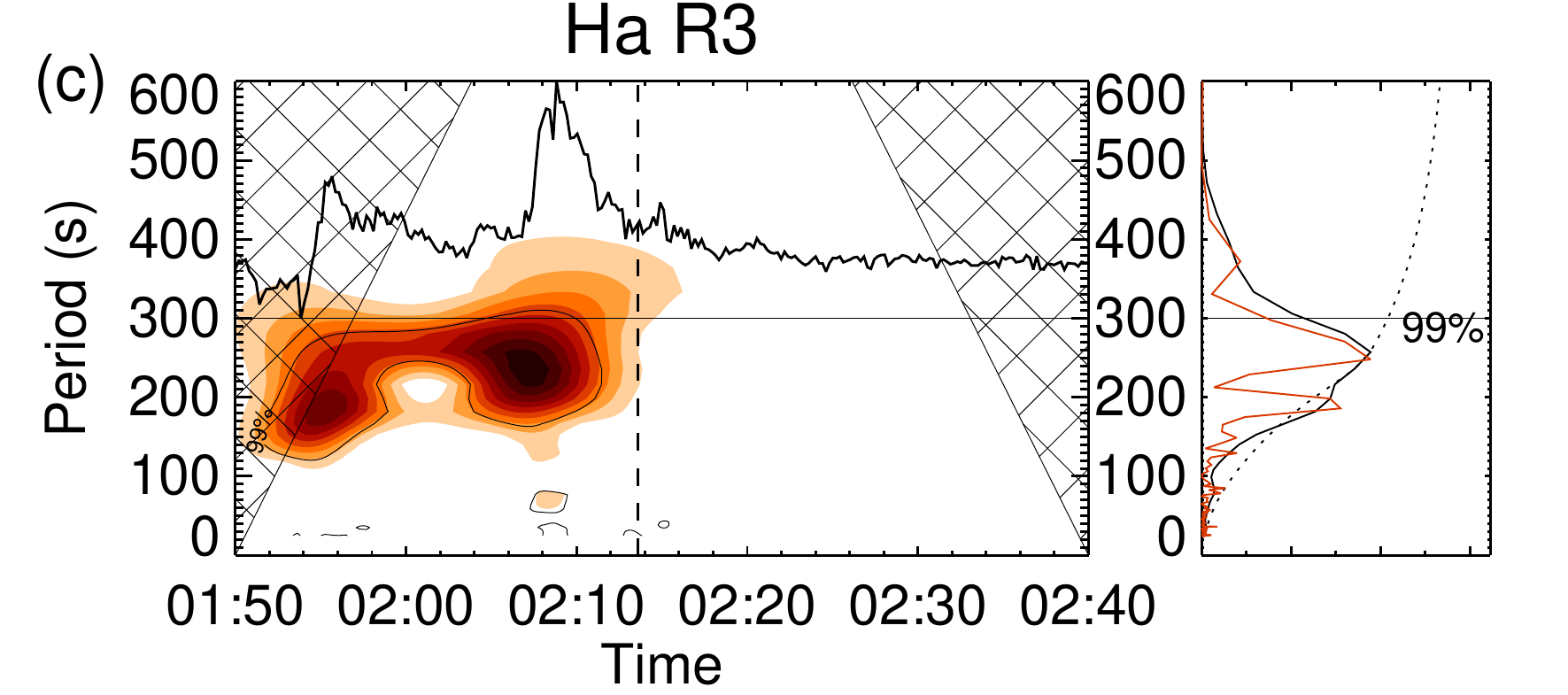}
	}
	\caption{Results of the wavelet and Fourier analyses of the  $H\alpha$ emission from different flare kernels. 
		The vertical dashed line in panel (d) indicates the start time of the brightening of the R4 ribbon, 02:13:20~UT. 
	}
	\label{f:waveletsHa}%
\end{figure*} 
\begin{figure*}
   \centerline{
\includegraphics[width=0.45\textwidth,clip=]{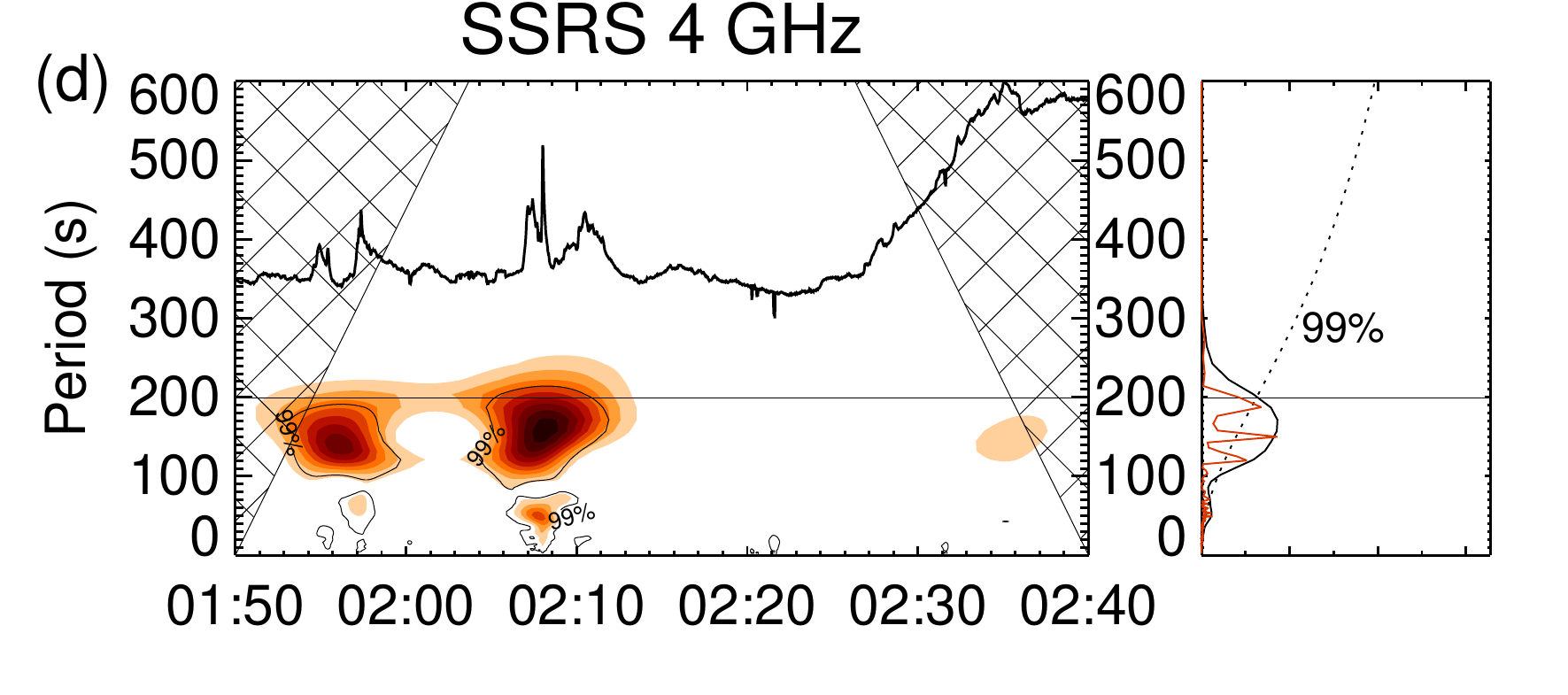}
\includegraphics[width=0.45\textwidth,clip=]{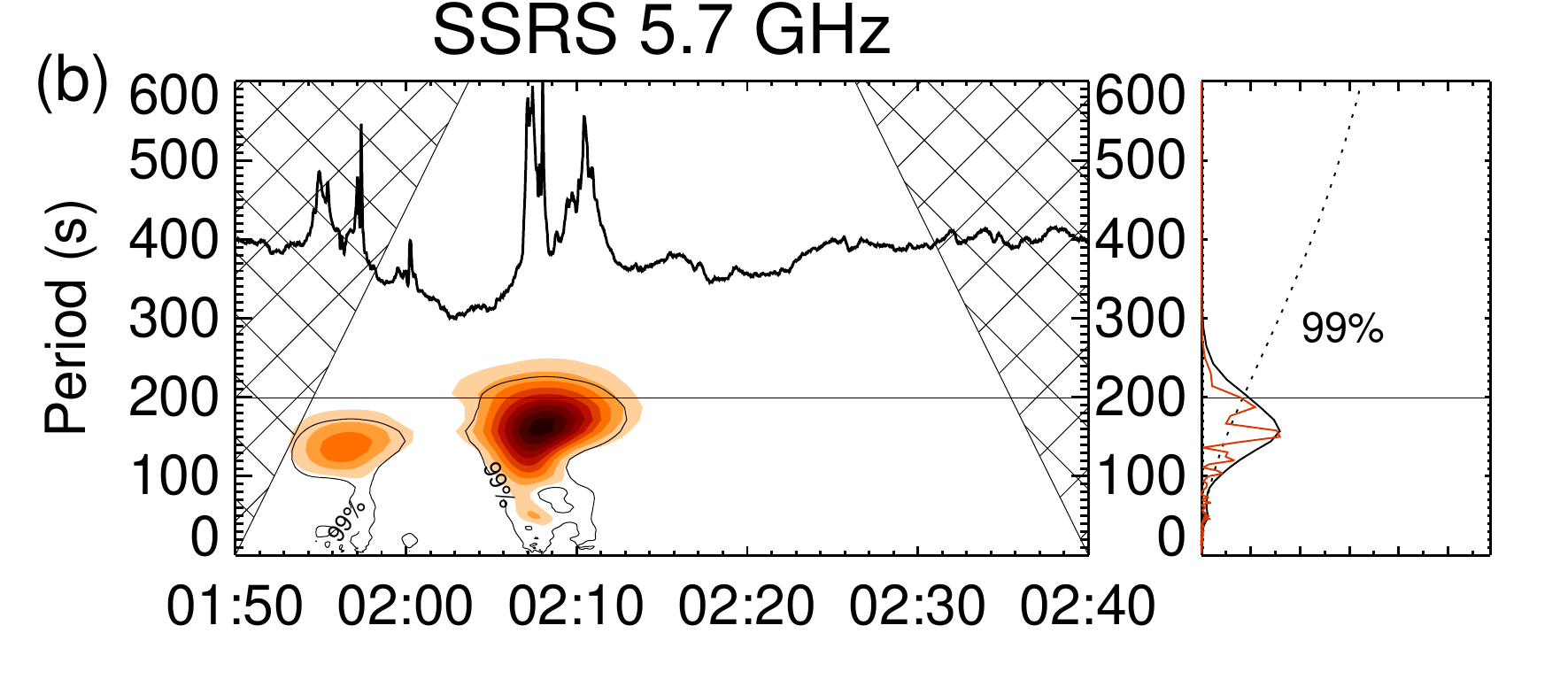}
   }
   \centerline{
\includegraphics[width=0.45\textwidth,clip=]{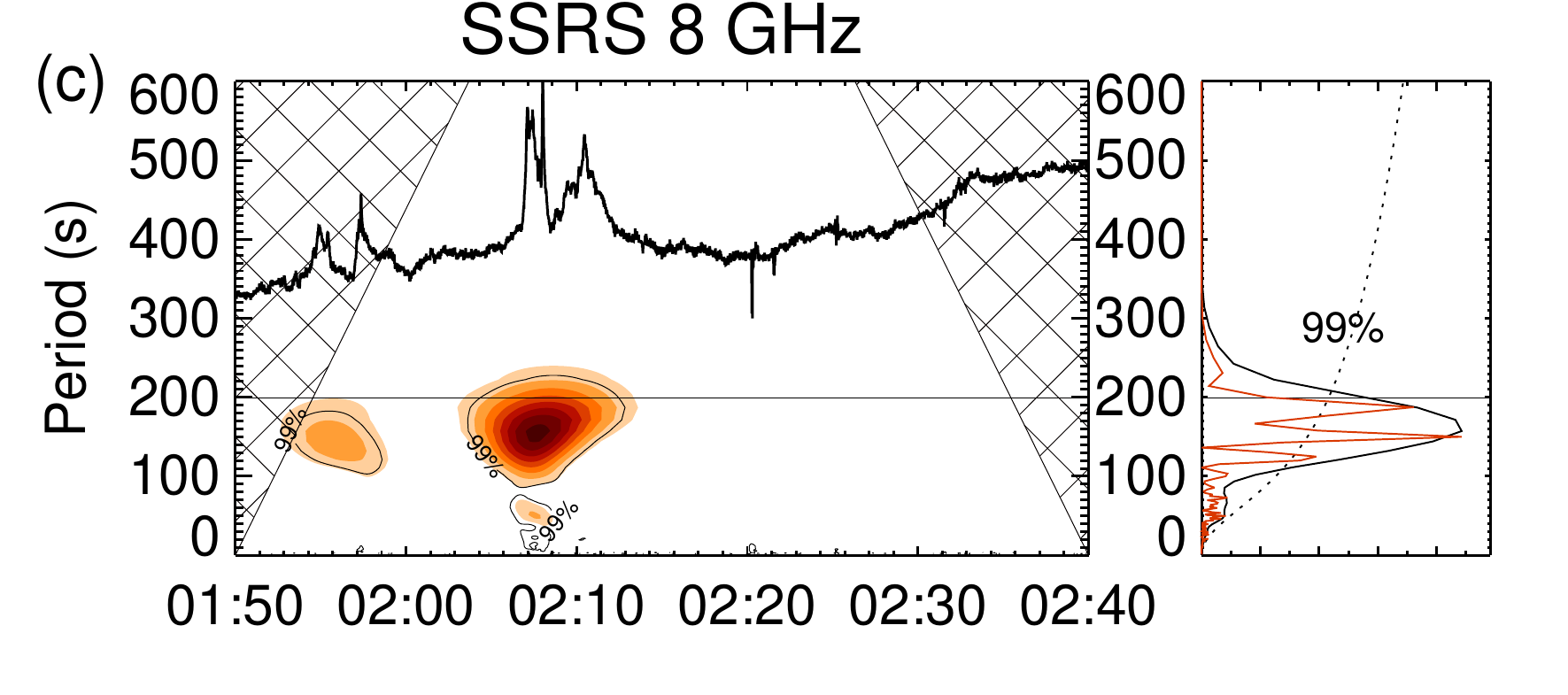}
\includegraphics[width=0.45\textwidth,clip=]{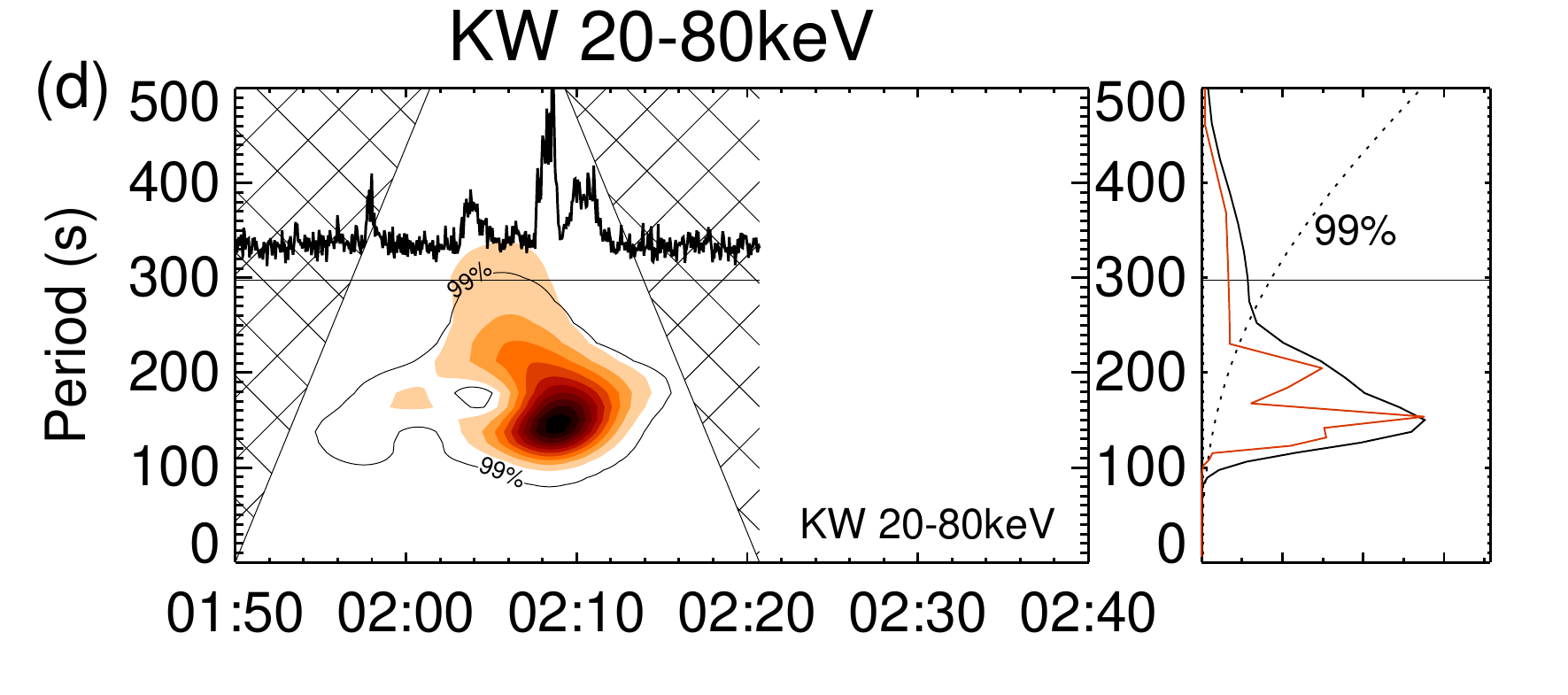}
   }
   \centerline{
\includegraphics[width=0.45\textwidth,clip=]{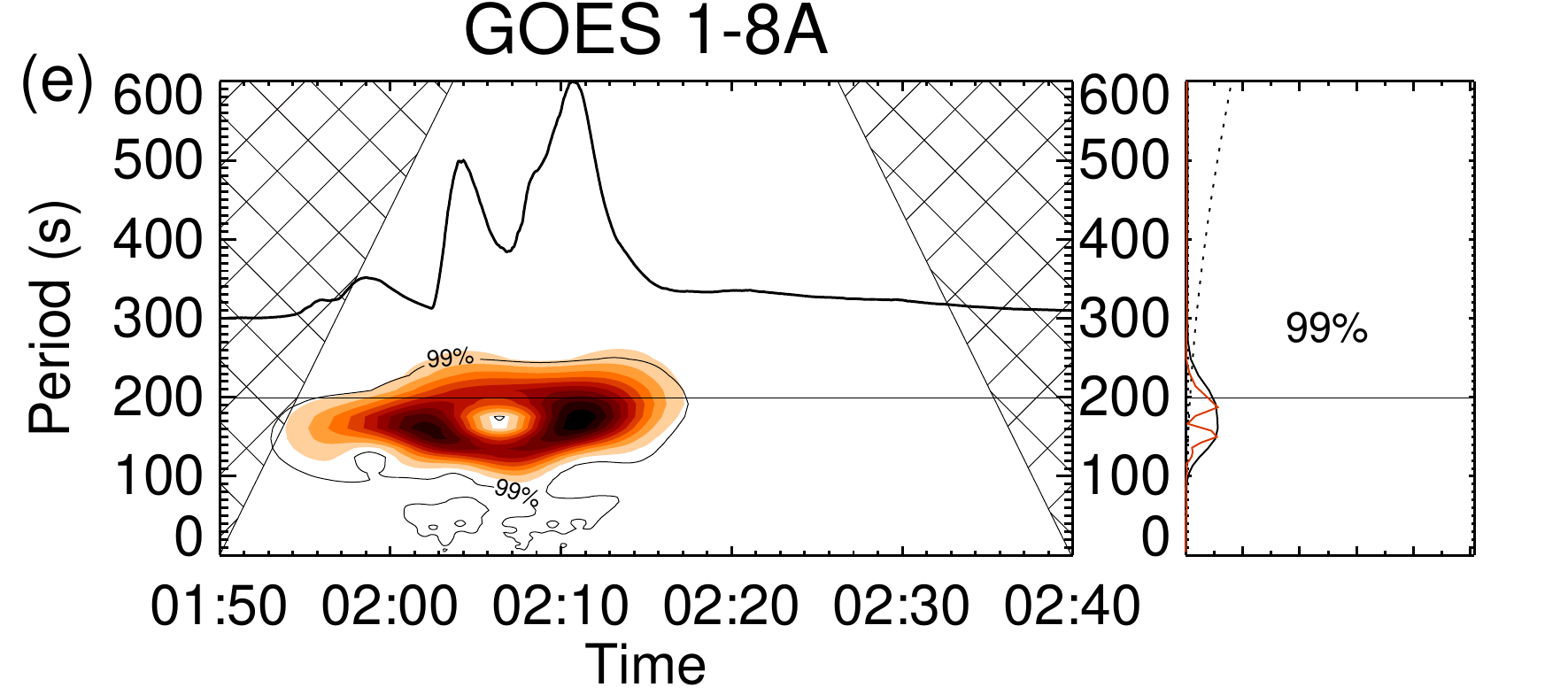}
\includegraphics[width=0.45\textwidth,clip=]{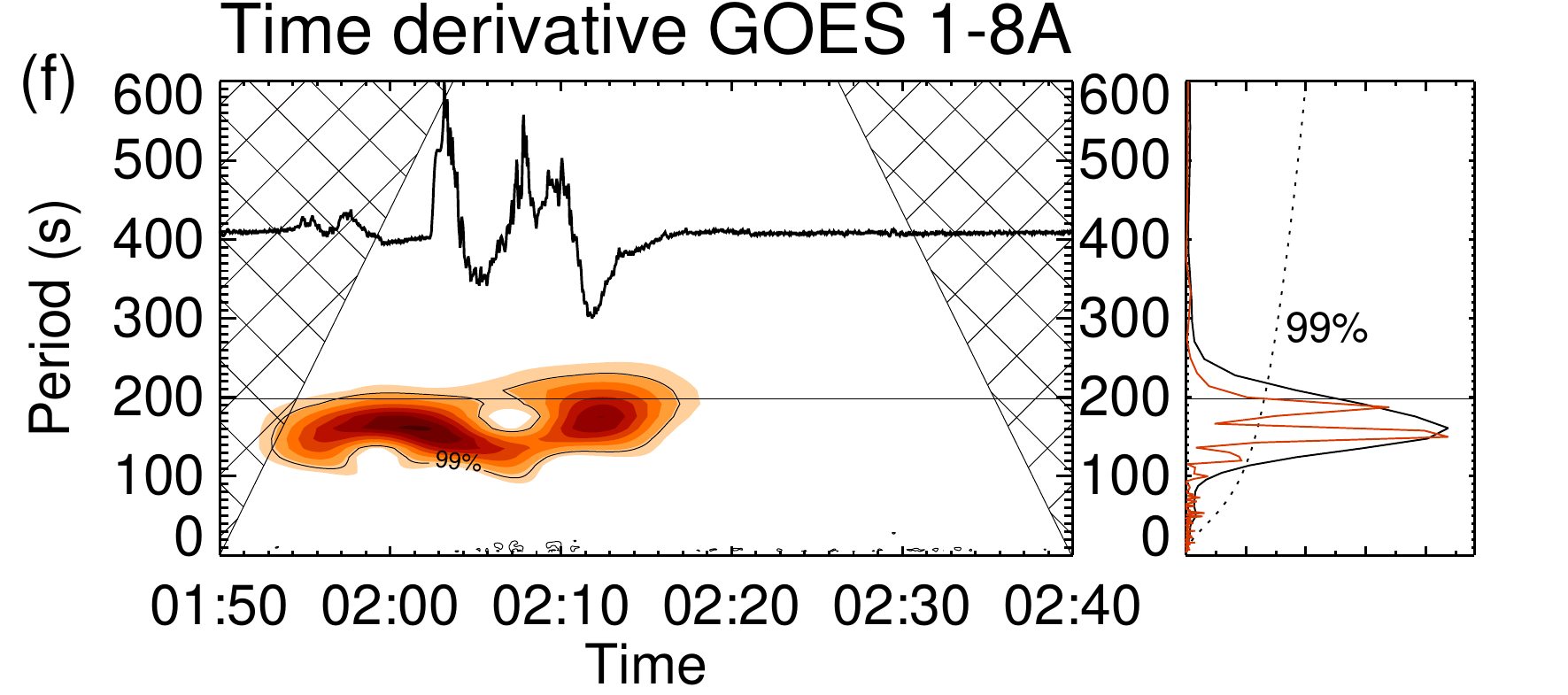}
   }

  \caption{Results of the wavelet and Fourier analyses of the microwave and X-ray emission. The vertical dashed line in panel (b) marks the start time of the brightening of the R4 ribbon. The vertical dashed line in panel (e) marks the start time of the interval analysed in Figure~\ref{f:waveletsMWXdecay} (panel (a)). 
}
  \label{f:waveletsMWX}%
\end{figure*} 

\begin{table*}
	\caption{Periods (in seconds) with error bars detected during both flares in different emission bands. The first two columns correspond to $H\alpha$ emission from the chromospheric kernels R1+R2 and R3, the next three columns correspond to the microwave emission at, correspondingly, 8~GHz, 5.7~GHz and 4~GHz registered with SSRS, the next two columns correspond to X-ray emission registered by KW and GOES, and the last column corresponds to time derivative from GOES flux.}              
	\label{t:periods_tot}      
	\centering                                      
	\begin{tabular}{ l  l  l  l  l  l  l  l }          
		\hline                        
		\multicolumn{2}{c}{$H\alpha$} & \multicolumn{3}{c}{SSRS} & \multicolumn{3}{c}{X-rays}  \\    
		R1+R2                    & R3                         & 8~GHz                & 5.7~GHz                 &        4~GHz           &  20--80~keV         & 1--8~{\AA} & 1--8~{\AA} ($dF/dt$) \\
		\hline                                   
		& $186^{+12}_{-11}$  & $187^{+13}_{-11}$ & $187^{+13}_{-11}$ & $187^{+13}_{-11}$ &     & $187^{+12}_{-11}$ & $187^{+12}_{-11}$  \\ 
		$148^{+8}_{-7}$    &      & $150^{+8}_{-7}$  & $150^{+8}_{-7}$    & $150^{+8}_{-7}$  & $153^{+13}_{-12}$ & $150^{+8}_{-7}$ &  $150^{+8}_{-7}$  \\
		\hline                                            
	\end{tabular}
\end{table*}
This section includes the results of analysis of QPPs  in the multi-wavelength emission of 
the series of the impulsive energy releases
occurred on March 5, 2014, between 01:50:00~UT and 02:40:00~UT. 
The details of the technique for the QPP analysis are presented in Appendix~\ref{s:Method}.
The results are collected in Figure~\ref{f:waveletsHa} and Figure~\ref{f:waveletsMWX}, in Table~\ref{t:periods_tot} for the impulsive flare phase and in Figure~\ref{f:waveletsMWXdecay} and Table~\ref{t:periods_t2} for the decay phase. 

The results for each time profile in Figures~\ref{f:waveletsHa}, \ref{f:waveletsMWX}, \ref{f:waveletsMWXdecay} are split into two sub-panels. The normalised time profile is plotted as a black curve in the left sub-panel. The time profile is normalised over its maximal amplitude, then is scaled according to the half maximum of the vertical axis and finally is shifted to the top of panel. The time profile is plotted over the red-colour gradient image which is the wavelet power spectrum. The right sub-panel shows the global wavelet spectrum (black solid curve) with its significance level (dashed curve). A Fourier periodogram normalised over the maximum of the global wavelet spectrum is plotted by the red curve. Both the wavelet Morlet transform and Fourier periodogram were applied to the high-frequency component which is residual after removing a slowly-varying flare trend from the time profile. We obtain the trend by a different methods but, in the figures, we show the results for a Savitzky-Golay polynomial filter having a width of $\tau = 200$~s (see Appendix~\ref{s:Method} for details of the data processing technique). The exceptions are panel (c) in Figure~\ref{f:waveletsHa} and panel (d) in Figure~\ref{f:waveletsMWX} where we selected $\tau = 300$~s to allow for longer periods. These $\tau$ values are shown in each wavelet spectrum by a horizontal line. It should be noted that smoothing a time profile over a $\tau$ value does not mean that the spectral powers in its wavelet of Fourier spectrum will be zero at the periods $P$ greater than the $\tau$ value. Their spectral powers will gradually tend to zero at longer periods. So, the peaks with $P > \tau$ will still be present in the spectrum but their powers will be suppressed. Notably the value of the true periods will not change when changing $\tau$. The right sub-panel collects two curves. The global wavelet spectrum obtained by integration of the wavelet power spectrum over time is shown by black solid curve.

We checked the significance level for two kinds of noise spectra and found that usage of the white noise leads to most  spectral peaks being significant, making the results uninformative. To the contrary, application of a red noise background spectrum leads to more reasonable constraints on the number of significant components. The 99\% significance level according the red noise is plotted as black thin contour over the wavelet spectrum and as the dashed line over the global wavelet spectrum. Fourier periodogram of the high-frequency signal is plotted by a red curve. The periodogram is normalised over the maximum of the global wavelet power.

The spectral resolution is not the same for longer and shorter periods in the periodogram because its period axis is defined as $P_i = 1/f_i$, where $f_i$ are the Fourier frequencies, $i = 0 .. N/2$, $N$ is the number of points in the time profile. We need to take into account that the period axis is not regular when estimating the error bars for periods. In the next subsections we will give the period values in the following format:  $P \approx A^{+er_1}_{-er_2}$~s. Here, $A = 1/f$ is the period (in seconds) corresponding to the Fourier frequency $f$ at the peak maximum. The superscript $er_1$ and subscript $er_2$ mean the differences between $A$ and the two neighbour periods in the period axis. As the period axis is not regular, the values $er_1 \neq er_2$. 

\subsection{Chromospheric $H\alpha$ emission}\label{s:PeriodsHa}
The periodic properties of $H\alpha$ emission from different sources are collected in Figure~\ref{f:waveletsHa}.  The periodic variations are pronounced in the time profiles associated with both the whole
 structure R1+R2 and the inner ribbon R1 (the thick black curve in the coloured panels (a) and (b)).  Wavelet spectra demonstrate well recognised stripes of stable period. Wavelet transform detects the period $P \approx 140^{+12}_{-11}$~s. At the same time, the Fourier periodogram reveals one dominant peak at the period $P \approx 148^{+8}_{-7}$~s. The peak in global wavelet spectrum (black curve) is wide.  Its full width at the half maximum is $FWHM \approx 60$~s.  The main peak in the periodogram (red curve) is narrower than the corresponding peak in the global wavelet spectrum, $FWHM \approx 10$~s.  It is obvious from the wavelet spectrum that the periodicity found is  associated with the flare activity.  No periodicity is evident before the flares nor after the flares. In the wavelet spectrum of the whole circular ribbon structure R1+R2 (Figure~\ref{f:waveletsHa}, panel (a)), the periodicity continues for a long time, 1400~s, enveloping nine period cycles in total. The value of the period does not depend on the value of the smoothing interval $\tau$ (see Section~\ref{s:Method} for details). Therefore, we show the results for $\tau = 200$~s in panels (a) and (b) for better visualisation.
 
The wavelet spectrum of the remote source R3 (Figure~\ref{f:waveletsHa}, panel (c)) reveals a number of irregular spots during the flares. The global wavelet spectrum shows a very wide peak with a maximum at $P \approx 152$~s with the $FWHM \approx 90$~s. A Fourier periodogram 
 demonstrates the set of three peaks with the periods $125$~s, $150$~s, and $185$~s. These peaks have almost equal powers. However, the periods $125$~s and $150$~s disappear for  $\tau = 300$~s leaving the only significant period  $186^{+12}_{-11}$~s. Therefore, we show the results for $\tau = 300$~s.

\subsection{Microwave and hard X-ray emissions}\label{s:PeriodsMWX}
In order to study in detail the periodicity found in $H\alpha$ emission from the circular ribbon structure during two flares, we make a comparative analysis of the periodicities in different wave bands. Multi-periodic behaviour is found both in microwave and hard X-ray emissions (Figure~\ref{f:waveletsMWX}). Each of the wavelet spectra of the microwave emission at 4~GHz, 5.7~GHz, and 8~GHz reveals a period of  $157$~s that coincide with the flares. Each of the global wavelet spectra shows wide peaks with maxima at the period $P \approx 157 ^{+14}_{-13}$~s.  

Fourier periodograms at these three frequencies show similar, but not identical, compositions of spectral components. The periodogram at 4~GHz shows three peaks with maxima at periods $187$~s, $150$~s, and $128$~s (Figure~\ref{f:waveletsMWX}, panel (a)). However, the periodogram applied to the filtered time series (with the Gaussian-shaped Fourier filter with boundaries 120~s and 200~s enveloping these three spectral peaks) leaves only two significant periods $187^{+13}_{-11}$~s, $150^{+8}_{-7}$~s. The global wavelet spectrum does not identify two periods, enveloping them with one broad peak ($FWHM \approx 68$~s) with maximum at $P \approx 172$~s. The wavelet transform identifies the period $P \approx 150$~s with each flare and the period $P \approx 187$~s with the interval between the flares. Also, we note that the period 187~s is comparable with the duration of the flares observed in microwaves. 
The spectral peak with the period $P \approx 128$~s has disappeared from the wavelet spectrum and from the periodogram. This indicates that this peak could be a harmonic of a low-frequency trend \citep{2009A&A...493..259I}. 

In the same way, the periodogram of 5.7~GHz emission contains peaks at periods of $187^{+13}_{-11}$~s and $150^{+8}_{-7}$~s  (Figure~\ref{f:waveletsMWX}, panel (b)). Emission at 8~GHz reveals the following set of the periods $187^{+6}_{-69}$~s, $150^{+4}_{-4}$~s, and $128^{+4}_{-4}$~s  (Figure~\ref{f:waveletsMWX}, panel (c)). However, the Fourier filtration method reveals only the first two periods.

The wavelet power spectrum for hard X-ray emission at 20--80~keV demonstrates complicated period structure from 01:56~UT to  02:15~UT (Figure~\ref{f:waveletsMWX}, panel (d)). The global wavelet spectrum shows one wide peak with maximum at $P \approx 150$~s. 
The overplotted periodogram shows that the peak consists of one dominant peak $P \approx 153^{+13}_{-12}$~s and two secondary, less powerful peaks. The periods of the secondary peaks shown in the figure are $P \approx 131^{+10}_{-8}$~s and $P \approx 204^{+25}_{-20}$~s. However, the period 204~s is comparable with the duration of the flares themselves, i.e. it is rather a trend than a QPP. The period 131~s is not stable relative to the $\tau$ value. Therefore, we do not consider them in the following analysis.

\subsection{Soft X-ray emissions}\label{s:PeriodsSHR}
The wavelet spectrum of the soft X-ray emission at 1--8~{\AA} and its time derivative are presented in Figure~\ref{f:waveletsMWX} (panels (e) and (f)). Testing by each of the methods mentioned in Section~\ref{s:Method} reveals two significant periodic components in the 1--8~{\AA} channel, $187 ^{+12}_{-11}$~s and $150^{+8}_{-7}$~s. The time derivative of the flux at 1--8~{\AA} contains the same periods. 

\subsection{Periods during the flare decay}\label{s:PeriodsR4}
\begin{figure*}
	\centerline{
		\includegraphics[width=0.45\textwidth,clip=]{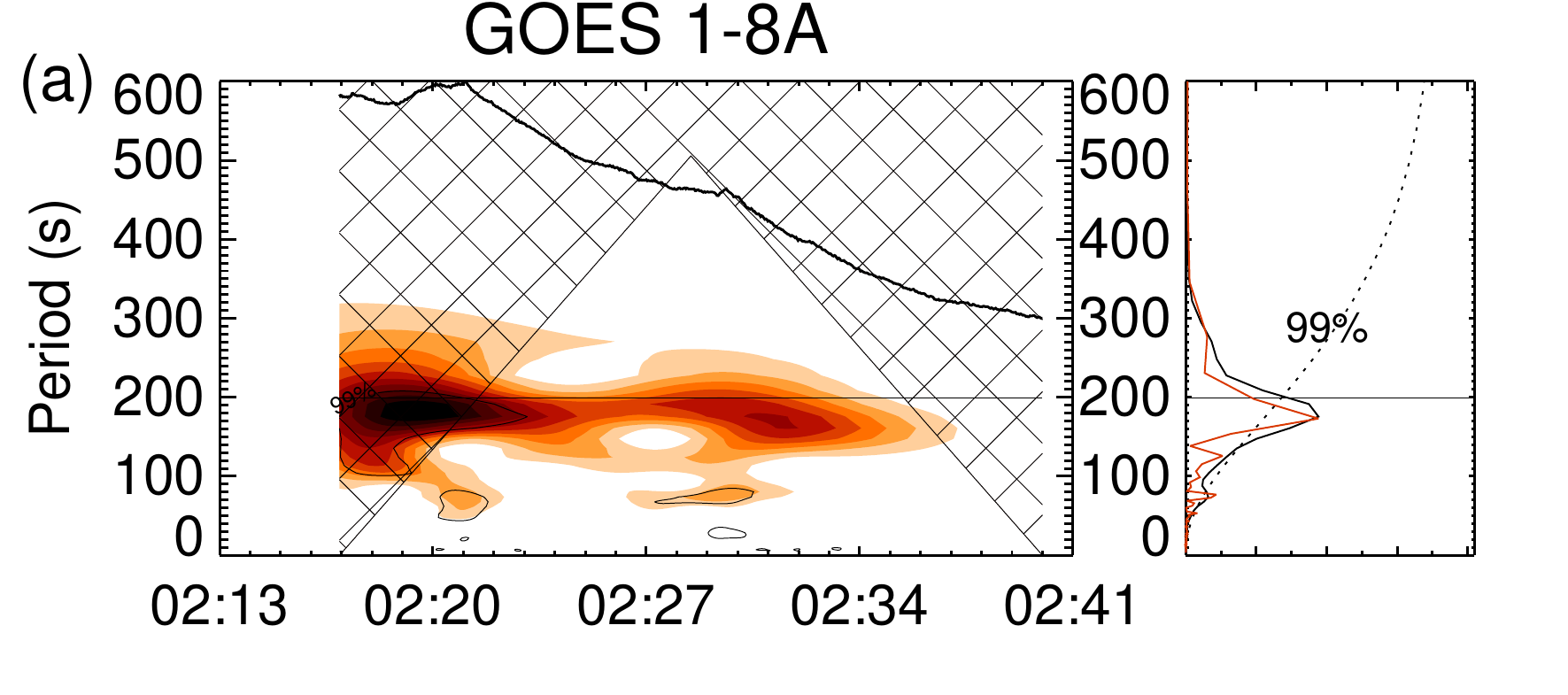}
		\includegraphics[width=0.45\textwidth,clip=]{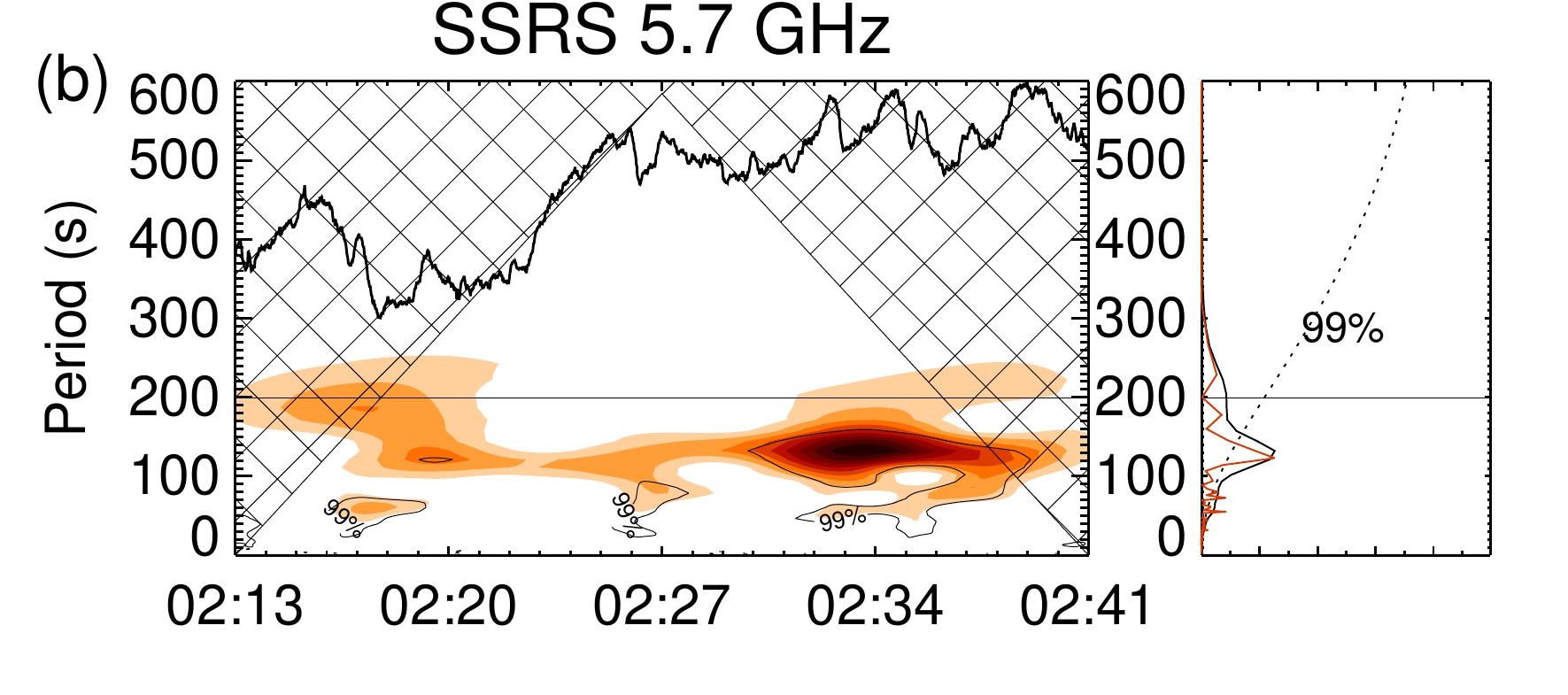}
	}
	\centerline{
		\includegraphics[width=0.45\textwidth,clip=]{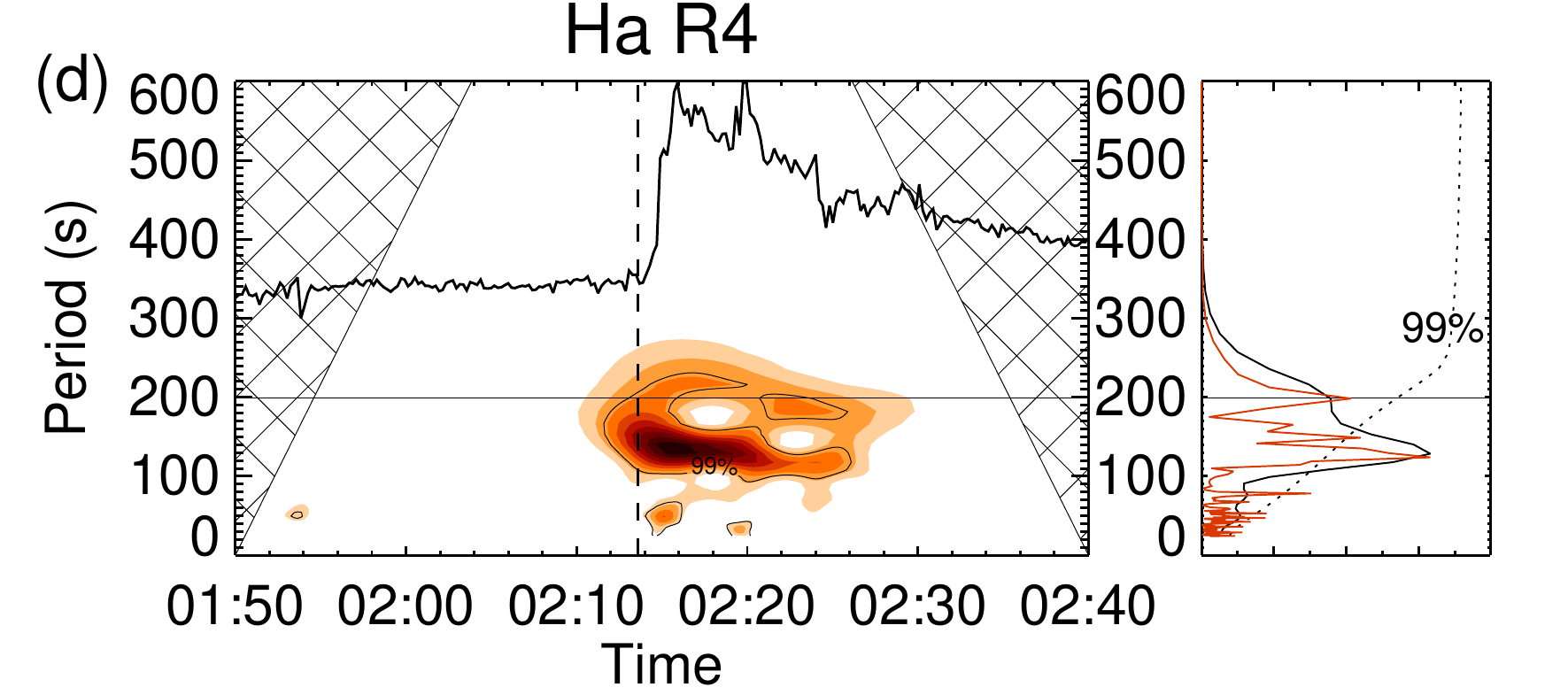}
		\includegraphics[width=0.45\textwidth,clip=]{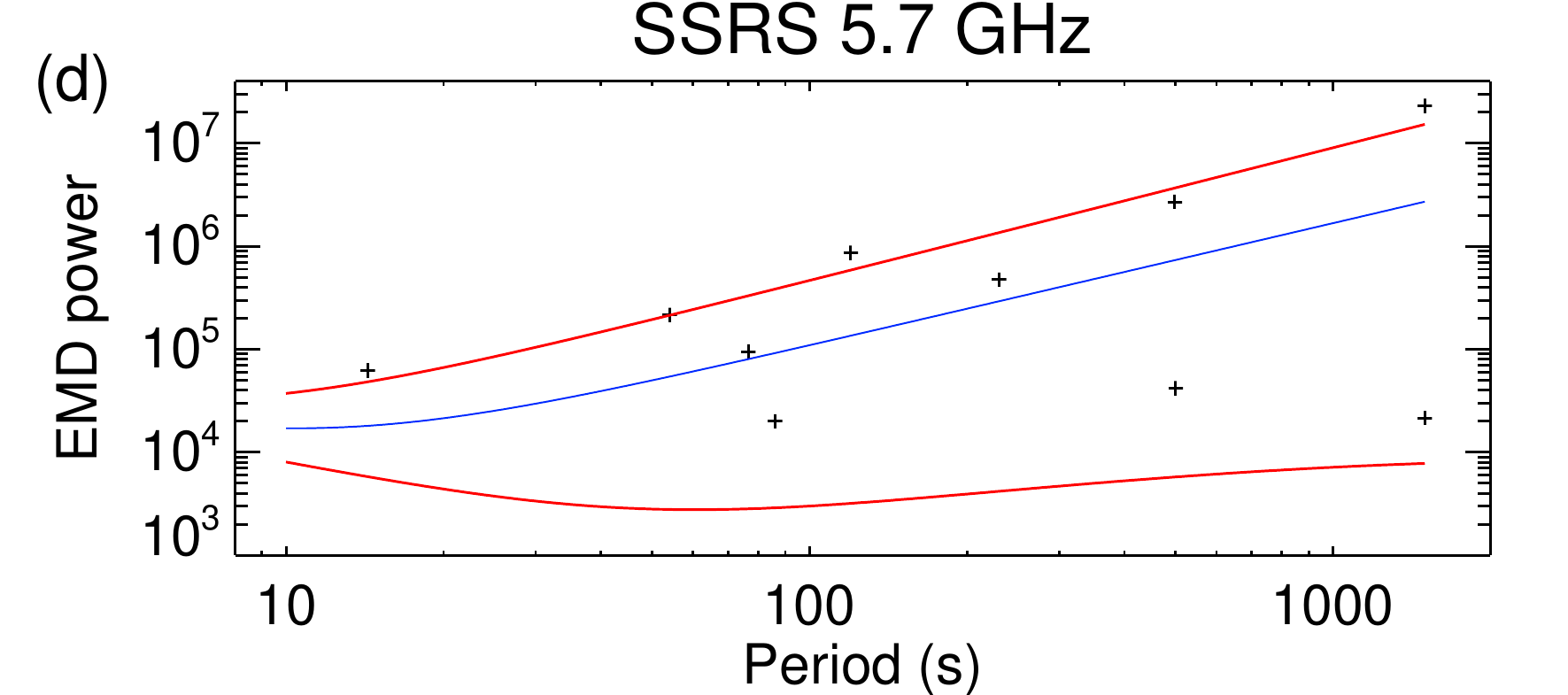}
	}
	\caption{Panels(a)--(c): results of the wavelet and Fourier analyses of the soft X-ray, microwave and $H\alpha$ R4 kernel data, during the decay flare phase.
	Panel (d): the empirical mode decomposition (EMD) spectrum of the microwave flux at 5.7~GHz. The total energies of the decomposed intrinsic modes are shown by crosses. The blue curve represents the global slope of the spectrum. The 99\% confidence intervals are shown by red curves.
	}
	\label{f:waveletsMWXdecay}%
\end{figure*} 
\begin{table}
	\caption{Periods (in seconds) with error bars detected during the decay phase  in different emission bands. The first column corresponds to $H\alpha$ emission from chromospheric R4 kernel, the second column corresponds to the soft X-ray emission registered with GOES, and the last column corresponds to the microwave emission at 5.7~GHz registered with SSRS.}              
	\label{t:periods_t2}      
	\centering                                      
	\begin{tabular}{ l  l  l }          
		\hline                        
		$H\alpha$          & GOES               & SSRS \\
		R4         & 1--8~{\AA}             & 5.7~GHz  \\
		\hline                                   
		$197^{+13}_{-12}$   & $187 ^{+13}_{-12}$  &                                \\ 
		$128^{+11}_{-11}$ &                                     & $123^{+10}_{-9}$  \\  
		\hline                                            
	\end{tabular}
\end{table}
The brightening of the elongated ribbon R4 started when the emission from the other $H\alpha$ ribbons and the emission in the other wavelengths reached the decay phase (Figure~\ref{f:timeprofs}). In contrast to the microwave emission, the soft X-ray emission still contains strong periodic components connected to the impulsive flare phase. Therefore, in order to avoid the influence of the periodicities related to the flares we analyse GOES data that started later than $H\alpha$ and microwave data. Those start times are marked both by dashed vertical line in Figure~\ref{f:waveletsHa} and Figure~\ref{f:waveletsMWX} (02:13:20~UT) for the microwave emission  and by dotted vertical line in Figure~\ref{f:waveletsMWX} (02:16:56~UT)~--- for the soft X-ray emission. The results are shown in Figure~\ref{f:waveletsMWXdecay} and in Table~\ref{t:periods_t2}. 

The global wavelet spectrum of ribbon R4 (Figure~\ref{f:waveletsHa} panel (d)) has a maximum at the period $P \approx 128^{+11}_{-11}$~s. This period keeps its value and significance for all the considered $\tau$ values. There is a secondary peak at $P \approx 197^{+13}_{-12}$~s. This period also keeps its value for $\tau > 200$~s. Note that the wavelet spectrum in  Figure~\ref{f:waveletsHa} (panel(d)) corresponds to $\tau = 200$~s to emphasise the shorter period, therefore the 197~s period is not significant there. The longer period is also found in GOES data. Its wavelet spectrum (Figure~\ref{f:waveletsMWXdecay} panel (a)) reveals one peak at $P \approx 187^{+13}_{-12}$~s.

The microwave emission at 5.7~GHz contains only a single mode QPPs with  $P \approx 123^{+10}_{-9}$~s (Figure~\ref{f:waveletsMWXdecay} panel (b)). However, when increasing $\tau$ up to 300~s the period of 200~s became significant while the 123~s period decreases its power down to below the significance level. In order to perform an independent check of these periods we applied the method of Empirical Mode Decomposition (EMD). This method decomposes a signal into intrinsic modes which are, in contrast to wavelet and Fourier methods, {\it a priori} not restricted to the harmonic functions or their wavelets (see Appendix~\ref{s:Method} for details). The EMD spectrum of the flux at 5.7~GHz is shown in Figure~\ref{f:waveletsMWXdecay} (panel(d)). Crosses indicate the total energies for all detected EMD modes. The general slope of the EMD spectrum is plotted by blue curve. For each characteristic time scale (or period), the spectral energy is distributed as a $\chi^2$ function with a number of degrees of freedom $k > 2$, resulting into the existence of two, upper and lower, confidence intervals \citep{2016A&A...592A.153K}. Thus, the significant EMD modes are those with the total energies outside the area bounded by these intervals.  In panel (d), the total energy of the 120~s EMD mode is seen to be well above the 99\% upper confidence level, coinciding with the 123~s period found in wavelet analysis panel (b). The period of 190~s was not detected in the EMD analysis at all. 
In contrast to the microwave emission at 5.7~GHz, the combination of the mentioned methods did not detect any significant periods at both 4~GHz and 8~GHz.

\vspace {0.3 cm}
To summarise, the quasi-periodic variations with a common average period of 150~s are found during the impulsive phase of the flares in the time profiles of $H\alpha$ emission from the circular ribbon structure, in the microwave emission at 4--8~GHz, in the hard X-ray emission at 20--80~keV, and in soft X-ray emission in 1--8~{\AA} band including its time derivative. The value of the period varies from 148~s to 153~s for different observational bands, but the variations do not exceed the detection errors (see Table~\ref{t:periods_tot}).  We also find a period around 190~s is present in H$\alpha$ emission from R3 kernel, in microwave emission at 4--8~GHz, and in soft X-ray emission at 1--8~{\AA}. During the decay phase we also detected the same period around 190~s, within the error bars: the period 187~s in soft X-rays and 197~s in  H$\alpha$ emission from kernel R4. The  period with an average value of 125~s is found during the decay flare phase in the emission of the $H\alpha$ R4 kernel and in the microwave emission at 5.7~GHz.

\section{Discussion and conclusions}\label{s:Discussion}
We chose the circular ribbon flare on March 5, 2014 because of its location near to the disk centre that allowed us to observe the flare ribbons related to a magnetic fan and the remote flare ribbons with minimal projection effects. Studying this event, we used QPP analysis for understanding relationship between the different flare kernels and processes in the circle ribbon flares. We were interested in revealing the flare kernel (or kernels) that were the source of the observed series of pulses and determining the pulse's generation mechanism (or mechanisms). 

Following the analysis of the time profiles performed in Section~\ref{s:Timeprofs}
we found a periodicity ($P_1$) around 150~s in the H$\alpha$ emission from flare kernels R1, R1+R2, microwave emission and the X-ray emission (panels (a)--(d) in Figure~\ref{f:waveletsMWX}). Thus we can conclude that it is the same periodicity related to R1+R2 kernels and appeared in both X-rays and microwaves. We note that this periodicity is also seen in the GOES time derivative that is used as an indicator of the accelerating processes according to the Neupert effect. The presence  of a 150~s periodicity in the hard X-ray emission (20--80~keV), the GOES time derivative and the 4--8~GHz microwave emission suggests the non-thermal origin of this periodicity. The presence of such a period in soft X-ray flux 1--8~{\AA} indicates that plasma heating were fully caused by the thermalisation of accelerated electrons. Most likely, it is the result of repetitive injections of accelerated electrons.  The presence of only this period in the emission from  R1+R2 flare kernels implies that the magnetic fan-spine  (Figure~\ref{f:waveletsHa})  is the source of the observed QPPs with a period around 150~s.

The explanation of the 150~s periodicity is likely related to the continuous rotation of the parasitic polarity in the vicinity of the future circular ribbon and apparent motion of the emission source in the circular ribbon R2 reported by \cite{Xu2017ApJ}.  Previous studies \citep{2006SoPh..238..347A,2009ApJ...700..559M,Ried2012A&A,2012ApJ...760..101W} noted the presence of slipping and  slip-running reconnection in the fan-spine, and related it with the apparent propagation of the circular ribbon emission.  \citet{2012ApJ...760..101W} found evidence of slipping or slip-running reconnection in their studied events, and the emission propagation was in the anticlockwise direction as reported by \cite{Xu2017ApJ} for the current event. 

We can do some estimations assuming that the length of the H$\alpha$ R2 ribbon during the initial brightening was the result of the primary energy release and precipitation of accelerated electrons. The diameter of the outline of the fan-spine is about 10~arcsec (see Figure~\ref{f:SDO}), giving the full length of the perimeter to be 23~Mm. 
As one can see in Figure~1 in \citet{Xu2017ApJ}, the initial R2 ribbon length is less than one-quarter of the outline length. Thus the ribbon length ($L_r$) is about 6~Mm.  We can assume that we observe the direct precipitation of electrons accelerated during the reconnection process within this ribbon length. In this case, the velocity of the process propagation should be up to $L_r/P_1$. So for $P_1 = 150$~s this gives $40~\rm{km~s}^{-1} $.
Estimations of characteristic slipping reconnection velocities from simulations were around $30~\rm{km~s}^{-1}$ \citep{2009ApJ...700..559M}.  Observational estimates vary within 20--40~$\rm{km~s}^{-1}$ \citep{2016ApJ...823...41D}. This agreement between our estimations and the velocity measurements indicate slipping or slipping running reconnection as the source of quasi-periodicity detected in the R1 and R2 kernels. 
The other possible drivers of reconnection processes are magnetohydrodynamic (MHD) waves. The sound speed~--- the characteristic phase speed of the slow magnetoacoustic
mode~--- estimated for the electron temperature 14--17~MK (see Section 4) is $C_s \approx$~620--680~$\rm{km~s}^{-1}$. If the reconnection is modulated by the slow MHD wave propagating along an arcade formed by the magnetic fan, the sound speed is an order of magnitude higher than the velocity of process propagation obtained above (about 40~$\rm{km~s}^{-1}$). The propagation of the fast MHD wave is defined by the phase speed of the fast MHD mode, which is higher than the sound speed. All these estimations of MHD wave characteristics do not support the hypothesis of an MHD wave as the direct reason of the observed 150~s periodicity in the case of  wave propagation along the circular arcade. However, we note that in the 3D case, the propagation parameters of MHD waves could change due to various reasons \citep[see][]{2016SoPh..291.3185A, 2011ApJ...730L..27N} and could potentially explain the observed 150~s periodicity. However, the in-depth analysis required is beyond of the scope of this study.

According to Figure~\ref{f:SDO} (bottom panel) and the 3D magnetic field reconstruction 
\citep[see Figure 8 in][]{Xu2017ApJ}, 
the remote flare kernel R3  was connected to the magnetic fan-spine with a loop. So, we could expect that a periodic process in the circular ribbon structure affects the periodic properties of the $H\alpha$ emission from the R3 kernel, for example, {\it via} accelerated electrons passing from the reconnection site to the R3 kernel along the loop. However, we did not detect any period around 150~s in the R3 kernel. The dominant period found in the $H\alpha$ emission from R3 kernel is 190~s. 
The 190-s period is absent in the hard X-ray emission that is related to non-thermal electron precipitation only. 
Moreover, there are obvious differences between the time profiles of the R1, R2 and R3 kernels and hard X-ray emission. These facts do not support the hypothesis of direct excitation of R3 emission by the same population of accelerated electrons that generated the emission of R1 and R2 kernels.

On the other hand, we found the 190~s period in the microwave and soft X-ray emission. The kernel R3 is the likely source of this period in flare emission as it is located very close to the large sunspot and possessed the same magnetic polarity as the sunspot. The value of the period is close to the 180~s period oscillation usually observed in sunspots~\citep[see, for example,][]{2008sust.book}. Thus, oscillations leaking from the sunspot could modulate both the R3 flare $H\alpha$ emission and the gyrosynchrotron emission of accelerated electrons seen in the microwaves.

Emission of the kernel R4 started during the decay phase of R1, R2 and R3 kernels, and there are no hard X-ray emission or response in soft X-ray time derivative. Two periods were detected in kernel R4. The first period 190~s was detected in the 
soft X-ray emission while the second period 125~s was found in microwave emission.
These facts could indicate different origins of detected oscillations.

The similarity of the 190~s QPPs seen both in the H$\alpha$ emission from kernel R4 and in the soft X-ray emission indicates that a thermal mechanism caused these QPPs. We can attribute the detected 190~s periodicity
to the  oscillations in the sunspot having characteristic timescales within 2--4~minutes \citep[see][]{2015LRSP...12....6K,2015A&A...577A..43S}, or the 3-minutes sunspot oscillations. However, those oscillations do not have enough energy to generate the flare emission directly. Therefore, the period 190~s could be a result of the modulation of the flare emission from kernel R4 by the 3-minute sunspot oscillations. 

The second period, 125~s, found in R4 H$\alpha$ emission is similar to that found in the microwave emission at 5.7~GHz, indicating a non-thermal mechanism.
The fact that this periodicity is detected in only one microwave frequency could be caused by the low flux densities at other frequencies during the decay phase.
We hypothesise that the 125~s period could relate to kink oscillations of the loop connecting the fan (R1+R2) with remote kernel (R3) and seen in EUV emission black (bottom panel in Figure~\ref{f:SDO}). The loop or outer spine became visible at about 02:00~UT. One can see its transverse oscillations at 171~{\AA} band. The kink oscillations of the loop could cause a null point reconnection \citep{2006A&A...452..343N, Ried2012A&A,2012ApJ...760..101W}  and generate the accelerated electrons. The kink-oscillations of the loop could result in interaction between the loops relating the kernel R4 with the sunspot.  We speculate that part of the electron flux could precipitates and generates the kernel R4 emission. Moreover, the interaction between the loops could be another possible origin of the electron acceleration.
	
In summary, we conclude the following about the mechanisms of emission generation based on periodicity properties analysis in the different flare kernels. 
We revealed 150~s periodicity simultaneously observed in the spatially resolved H$\alpha$ flare kernels, soft and hard X-ray emission, and microwave emission. The H$\alpha$ flare kernels showing this periodicity were related with a magnetic fan-spine structure. The analysis of estimated parameters of the pulses suggests the slipping reconnection mechanism as the origin of these QPPs. 
We found the 190~s period in both the remote kernel R3 during the impulsive phase and the kernel R4 raised during the decay phase of the event. This periodicity is close to the 3-minute oscillations observing in sunspots. However, as the 3-minute oscillation usually do not have enough energy to directly generate the flare emission, we suggest that the QPPs of the flare kernels R3 and R4 could be caused by modulation of their emission by the 3-minute wave trains coming from the sunspot. 
The observational facts indicate a relation between the 125~s period found in the thermal kernel R4 emission and the kink oscillations of the outer spine which connects the place of the null point reconnection with the remote kernel, but this hypothesis needs additional study.
\begin{acknowledgements}
This research was partly supported by  the budgetary funding of Basic Research programs No. II.16 (LKK), No. II.12 (DYK) and No. 0041-2019-0019 (EGK), the grant of the Russian Foundation for Basic Research No. 18-02-00856 (EGK), the grant of the National Natural Science Foundation of China No.~11873091 and Basic Research Program of Yunnan Province No.~2019FA001 (ZX), by the STFC consolidated grant ST/P000533/1 (HASR), and by the STFC consolidated grants ST/P000320/1 and ST/T000252/1 (DYK).
Authors thank CSUC 'Angara' for SSRS data, teams operating RHESSI, FERMI, SDO, RSTN and GOES  for open access to observational data.
\end{acknowledgements}

\bibliographystyle{aa} 

\begin{appendix}
	\section{Time series processing method}\label{s:Method}
	The procedure described below is applied to each time series under study in a uniform way. The time profiles, first, are interpolated to a regular time axes. We consider different methods of data processing in order to minimise possible effects of the method on the results. We perform the periodic analysis with the de-trended time series, or high-frequency signal $F_{HF}$ \citep[see ][]{0741-3335-61-1-014024}. We use a combination of the wavelet (Morlet) technique \citep{1998BAMS...79...61T} and of the Lomb-Scargle periodogram \citep{1982ApJ...263..835S} and autocorrelation analysis. The method was tested and described in details by \citet{2010SoPh..267..329K}. We will mention its principal steps. 
	
	We attract both a smoothing methods and Fourier filtration method to obtain the high-frequency signal. The smoothing method implies, first, calculation of the slowly varying background, or trend, of the original time series. We use both smoothing with the running average (\texttt{smooth.pro}) and a Savitzky-Golay polynomial filter (\texttt{savgol.pro}), both IDL procedures are applied for the same characteristic time window $\tau$. Then, subtracting  the trend from the original time series, we obtain the high-frequency component. 
	 Method of  Fourier filtration implies zeroing the spectral power in the Fourier power spectrum of the original time series outside the period band (filter) of interest. The filter boundaries define the lower and higher period values at the half power of the Gaussian bell. So, the periods lower than and higher than the filter are not cut off abruptly but smoothly decrease their specral power down to zero.  Performing inverse Fourier transform to the filtered spectrum, we obtain the high-frequency component without trend and without high-frequency noise. The de-trended components obtained by these three ways are ready for following period analysis. 
	
	We use a combination of the wavelet technique and of the Lomb-Scargle periodogram to analyse the periods in flare emission. By both methods, the significant spectral peaks are found. The significance level of a spectral peak is estimated assuming Markov red noise (or Brownian noise) \citep{1998BAMS...79...61T}. Note that the red noise imposes stronger restrictions for significance of a spectral peak than white noise \citep{2011A&A...533A..61G}. Those peaks which spectral power exceed 99\% probability level by two methods are believed to be true.  
	
This procedure is repeated for each set of iterations. The reason for the iterations is that the smoothing could lead to an appearance of false peaks in the periodogram and in the wavelet spectrum. 
In order to avoid these possible artefacts we obtain the high-frequency component for a set of various $\tau$ values in a wide range, from $\tau =50$~s to $\tau = 300$~s. 
Changing the $\tau$ value affects the statistical significance of the peak in the periodogram. However, it does not change the value of the true period \citep[see, for example, periodograms in ][]{2013SoPh..284..559K}. Therefore, we believe that the period is true if it satisfies the following conditions: (1) the period value does not depend on the specific $\tau$ value; (2) the peak remains significant for different $\tau$ values. 
We do not make checks for longer $\tau$ values because in our specific case the characteristic duration of separate flares themselves is, roughly, 300--400~s. The flare repeatability rate is a topic of a separate study, but in this paper we are interested in the fine time structure of the flare emission. 
	
Independently, Gaussian filtration over periods from 120~s to 200~s is applied to verify shorter periods and from 120~s to 300~s to verify longer periods. 
The visual control of similarity of the high-frequency signals obtained with both smoothing and filtration is performed, and the periods pronounced for all the methods are selected for the following periodic analysis.
	
Finally, the values of the significant periods obtained with both the wavelet and Fourier methods applied to all the high-frequency signals are compared. Those periods which are similar for all the methods are selected for the diagnostics of processes in the flare volume. 
	
Separately, in order to check the periodic properties of the emissions during the decay phase, we performed the additional analysis using the method of the Empirical Mode Decomposition (or EMD). The fundamental difference of this method from the methods based on the Fourier transform described above is that it does not imply {\it a priori} any basic functions \citep{2008RvGeo..46.2006H}. The method analyses the time scales of the time series itself, collecting the similar time scales into an intrinsic empirical mode. The iteration method distinguishes modes of different time scales, from the low-frequency trend to high-frequency noise. Similar to the methods based on the Fourier transform, not all intrinsic EMD modes necessarily represent a statistically significant oscillatory process \citep{2016A&A...592A.153K}. Therefore, in this study, we estimate the significance of EMD modes by approximating the shape of the EMD spectrum by a power law function and constructing confidence intervals.
	
\end{appendix}

\end{document}